\documentclass[11pt]{article}
\usepackage[utf8]{inputenc}
\usepackage{amsmath,amsfonts,amssymb}
\usepackage{amsthm}
\usepackage{latexsym}
\usepackage[noadjust]{cite}
\usepackage{graphicx}
\usepackage{xcolor}
\usepackage{dirtytalk}
\usepackage{mathtools}
\usepackage[margin=1in]{geometry}
\usepackage{thm-restate}
\usepackage{enumitem}
\usepackage{comment}
\usepackage[linesnumbered,ruled]{algorithm2e}
\usepackage[colorlinks=true, allcolors=blue]{hyperref}
\usepackage[nameinlink,capitalise]{cleveref}
\hypersetup{
    citecolor={violet}
}

\newtheorem{theorem}{Theorem}[section]
\newtheorem{corollary}[theorem]{Corollary}
\newtheorem{lemma}[theorem]{Lemma}

\newtheorem{claim}[theorem]{Claim}
\theoremstyle{definition}
\newtheorem{definition}[theorem]{Definition}
\newtheorem{remark}[theorem]{Remark}

\newcommand{\N}{\mathbb{N}}
\newcommand{\E}{\mathbb{E}}

\newcommand{\zo}{\{0, 1\}}

\newcommand{\eps}{\varepsilon}

\newcommand{\polylog}{\mathrm{polylog}}
\newcommand{\poly}{\mathrm{poly}}

\newif\ifdraft
\draftfalse

\newif\ifnotanon
\notanonfalse

\pdfoutput=1

\title{Characterizing the Distinguishability of Product Distributions through Multicalibration}
\author{Cassandra Marcussen\thanks{Supported in part by an NDSEG fellowship, and by NSF Award 2152413 and a Simons Investigator Award to Madhu Sudan.}\\Harvard University\\ \texttt{cmarcussen@g.harvard.edu}\and Aaron (Louie) Putterman\thanks{Supported in part by the Simons Investigator Awards of Madhu Sudan and Salil Vadhan and NSF Award CCF 2152413.}\\Harvard University\\ \texttt{aputterman@g.harvard.edu}\\ \and Salil Vadhan\thanks{Supported by a Simons Investigator Award.}\\Harvard University\\ \texttt{salil\_vadhan@harvard.edu}\\}

\date{\today}
\date{}

\begin{document}

\maketitle

\begin{abstract}
Given a sequence of samples $x_1, \dots , x_k$ promised to be drawn from one of two distributions $X_0, X_1$, a well-studied problem in statistics is to decide $\textit{which}$ distribution the samples are from. Information theoretically, the maximum advantage in distinguishing the two distributions given $k$ samples is captured by the total variation distance between $X_0^{\otimes k}$ and $X_1^{\otimes k}$. However, when we restrict our attention to $\textit{efficient distinguishers}$ (i.e., small circuits) of these two distributions, exactly characterizing the ability to distinguish $X_0^{\otimes k}$ and $X_1^{\otimes k}$ is more involved and less understood.

In this work, we give a general way to reduce bounds on the computational indistinguishability of $X_0$ and $X_1$ to bounds on the $\textit{information-theoretic}$ indistinguishability of some specific, related variables $\widetilde{X}_0$ and $\widetilde{X}_1$. As a consequence, we prove a new, tight characterization of the number of samples $k$ needed to efficiently distinguish $X_0^{\otimes k}$ and $X_1^{\otimes k}$ with constant advantage as 
    \[
    k = \Theta\left(d_H^{-2}\left(\widetilde{X}_0, \widetilde{X}_1\right)\right),
    \]
which is the inverse of the squared Hellinger distance $d_H$ between two distributions $\widetilde{X}_0$ and $\widetilde{X}_1$ that are computationally indistinguishable from $X_0$ and $X_1$. Likewise, our framework can be used to re-derive a result of Halevi and Rabin (TCC 2008) and Geier (TCC 2022), proving nearly-tight bounds on how computational indistinguishability scales with the number of samples for arbitrary product distributions.
    
At the heart of our work is the use of the Multicalibration Theorem (H\'ebert-Johnson, Kim, Reingold, Rothblum 2018) in a way inspired by recent work of Casacuberta, Dwork, and Vadhan (STOC 2024). Multicalibration allows us to relate the computational indistinguishability of $X_0, X_1$ to the statistical indistinguishability of $\widetilde{X}_0, \widetilde{X}_1$ (for lower bounds on $k$) and construct explicit circuits to distinguish between $\widetilde{X}_0, \widetilde{X}_1$ and consequently $X_0, X_1$ (for upper bounds on $k$).
\end{abstract}

\thispagestyle{empty}
\newpage

\tableofcontents
\thispagestyle{empty}
\addtocounter{page}{-2}
\pagebreak

\section{Introduction}

Given a sequence of samples $x_1, \dots , x_k$ promised to be drawn from one of two distributions $X_0, X_1$, each defined on a domain $\mathcal{X}$, a well-studied problem in statistics is to decide \emph{which} distribution the samples are from. 
Information theoretically, the maximum advantage in distinguishing $X_0, X_1$ given $k$ samples is $d_{\mathrm{TV}}(X_0^{\otimes k}, X_1^{\otimes k})$, where $X_b^{\otimes k}$ denotes the result of independently sampling $k$ times from $X_b$, and $d_{\mathrm{TV}}(p,q) =\frac{1}{2} \sum_{x \in \mathcal{X}} |p(x) - q(x)|$ denotes the total variation distance. This advantage can then be related to the single-sample relation between $X_0, X_1$ in a few ways:

\begin{enumerate}
    \item An upper bound is given by the inequality: 
    \begin{equation}
        d_{\mathrm{TV}}(X_0^{\otimes k}, X_1^{\otimes k}) \leq 1 - (1-d_{\mathrm{TV}}(X_0, X_1))^k \leq k \cdot d_{\mathrm{TV}}(X_0, X_1). \label{ineq:TV}
    \end{equation}
    \item To obtain a good 2-sided bound, we can use the \emph{Hellinger distance}: 
    \begin{equation} 
    1 - e^{-k d^2_{\mathrm{H}}(X_0, X_1)} \leq d_{\mathrm{TV}}(X_0^{\otimes k}, X_1^{\otimes k}) \leq \sqrt{2k d^2_{\mathrm{H}}(X_0, X_1)},
    \label{ineq:hellinger}
    \end{equation}
    where the Hellinger distance is defined as \[d^2_H(X_0, X_1) = \frac{1}{2}\sum_{i\in\mathcal{X}} \left(\sqrt{\Pr[X_0 = i]} - \sqrt{\Pr[X_1 = i]}\right)^2.\]
\end{enumerate}

In particular, Inequality~(\ref{ineq:hellinger}) above shows that $k = \Theta\left(1/d^{2}_H(X_0, X_1)\right)$ samples is both necessary \emph{and} sufficient for distinguishing $X_0, X_1$ with constant advantage. In contrast, observe that \cref{ineq:TV} in the first point above is actually \emph{not} tight. To illustrate, we can consider the two distributions below:

\begin{enumerate}
    \item $X_0$ such that $\Pr[X_0 = 1] = 1/2, \Pr[X_0 = 0] = 1/2$.
    \item $X_1$ such that $\Pr[X_1 = 1] = 1/2 + \eps, \Pr[X_1 = 0] = 1/2 - \eps$.
\end{enumerate}

\cref{ineq:TV} implies only that $d_{\mathrm{TV}}(X_0^{\otimes k}, X_1^{\otimes k})  \leq k \eps$, and thus does not rule out the possibility of distintiguishing $X_0, X_1$ with constant advantage after $k = O(1 / \eps)$ samples. However, as we see in the \emph{Hellinger}-distance formulation of \cref{ineq:hellinger}, in this example $k = \Theta(1 / \eps^2)$ samples is both necessary and sufficient to distinguish $X_0, X_1$ with constant advantage. Thus, a key advantage in using Hellinger distance to characterize statistical distinguishability is in getting \emph{instance-optimal} characterizations of the statistical distinguishability \emph{for every} pair of random variables. 

In computer science, this task of distinguishing distributions arises in many domains, from cryptography to computational complexity. However, we typically work with the computational analogue of total variation distance known as \emph{computational indistinguishability}. We say that $X_0, X_1$ are \emph{$\delta$-indistinguishable by circuits of size $s$} if for every function $f: \mathcal{X} \rightarrow \zo$ computable by size $s$ circuits, $|\sum_{x \in \mathcal{X}} f(x) \cdot (\Pr[X_0 = x] - \Pr[X_1 = x])| \leq \delta$. In this paper, we state most of our results using this concrete security formulation, where $s$ and $\delta$ are separate parameters. However, our results can be translated into the traditional asymptotic complexity formulation where $s = n^{\omega(1)}$, and $\delta = n^{-\omega(1)}$ for a security parameter $n$ and unspecified super-polynomial functions $n^{\omega(1)}$.
In fact, we are interested in the setting of {\em weak indistinguishability} where $\delta$ is non-negligible ($\delta \geq 1 / \poly(n)$) while $s$ remains super-polynomial ($s \geq n^{\omega(1)}$). We elaborate on the asymptotic formulations of our results in \cref{sec:asymptoticComplexity}.

The classic hybrid argument of Goldwasser and Micali \cite{GM84} shows that if $X_0, X_1$ are $\delta$-indistinguishable for circuits of size $s$, then $X_0^{\otimes k}, X_1^{\otimes k}$ are $k \delta$-indistinguishable for circuits of size $s$, which corresponds to the weaker bound in \cref{ineq:TV} above. However, a more refined understanding of (in)distinguishability under repeated samples has been elusive. Indeed, the work of Halevi and Rabin \cite{HR08} was the first to show that $X_0^{\otimes k}, X_1^{\otimes k}$ are $(1 - (1 - \delta)^k + \eps)$-indistinguishable for circuits of size $s' = s / \poly(k, 1 / \eps)$ (with subsequent improvements to the parameters being made by \cite{geier2022tight}). Note that when $s = n^{\omega(1)}$ and $k \leq \poly(n)$, we can take $\eps = n^{-\omega(1)}$ and retain $s' = n^{\omega(1)}$, so this bound agrees with the tighter bound in \cref{ineq:TV} above up to an additive negligible term.

This bound on computational distinguishability can be viewed as an extension of \cref{ineq:TV} to the computational setting. Similarly to \cref{ineq:TV} from the information-theoretic setting,  the characterization of \cite{HR08, geier2022tight} is \emph{not} an instance-optimal characterization of the computational distinguishability of random variables, and a computational analog of \cref{ineq:hellinger} is still missing. In this work, we overcome this shortcoming by building on the key technique of \emph{multicalibration}, as we explain below.

\subsection{Main Results}

In this work, we give a general way to reduce reasoning about computational indistinguishability to reasoning about the information-theoretic case. This also allows us to deduce as corollaries both the characterization achieved by \cite{HR08, geier2022tight} and a computational analogue of Inequality~(\ref{ineq:hellinger}) (with the total variation distance of $X_0, X_1$ replaced by the computational indistinguishability of $X_0, X_1$ and the Hellinger distance between $X_0, X_1$ replaced by a quantity we call the \textit{pseudo-Hellinger distance} between $X_0, X_1$).

\begin{theorem}\label{thm:main}
    For every pair of random variables $X_0, X_1$ over $\mathcal{X}$, every integer $s$ and every $\eps > 0$, there exist random variables $\widetilde{X}_0, \widetilde{X}_1$ such that for every $k>0$,
    \begin{enumerate}
        \item $X_b$ is $\eps$-indistinguishable from $\widetilde{X}_b$ by circuits of size $s$, for each $b\in \zo$.  \label{itm:indist-from-tilde}
        \item $X_0^{\otimes k}$ and $X_1^{\otimes k}$ are $(d_{TV}(\widetilde{X}_0^{\otimes k}, \widetilde{X}_1^{\otimes k})+2k \cdot \eps)$-indistinguishable by circuits of size $s$. \label{itm:indist-multi}
        \item $X_0^{\otimes k}$ and $X_1^{\otimes k}$ are $(d_{TV}( \widetilde{X}_0^{\otimes k}, \widetilde{X}_1^{\otimes k})- 2k \cdot \eps)$-distinguishable by circuits of size $s' =  O(sk/\eps^6) + \text{poly}(k/\eps)$. \label{itm:dist-multi}
        \item The statements above also hold with $\widetilde{X}_1 = X_1$, but $s' = O(sk/\eps^{12}) + \text{poly}(k/\eps)$.
    \end{enumerate}
\end{theorem}

\begin{remark}
    More generally, we can define indistinguishability with respect to an arbitrary class of functions $\mathcal{F}$. See \cref{sec:mainLB} and \cref{sec:efficient} for generalized versions of the different parts of the above statement.
\end{remark}

\begin{remark}
    We can also generalize the above theorem to arbitrary product distributions. We prove this version in \cref{section:general-product-distributions}.
\end{remark}

As elaborated upon in \cref{sec:overview}, the above theorem follows from assigning $\widetilde{X}_0$ and $\widetilde{X}_1$ to be a careful ``mixture'' of $X_0$ and $X_1$. This mixture depends on carefully partitioning the domain space of the random variables using \emph{multicalibration}. With these random variables, Item~\ref{itm:indist-multi} follows directly from Item~\ref{itm:indist-from-tilde} via hybrid argument, so the value of \cref{thm:main} is that it provides not only an upper bound on (in)distinguishability but a matching lower bound, via Item~\ref{itm:dist-multi}.
Thinking of $s = n^{\omega(1)}, k \leq \poly(n), \eps = 1/ s$, we see that up to a negligible additive term $(k \eps)$, and a polynomial change in circuit size ($s'$ vs. $s$), the computational indistinguishability of $X_0, X_1$ under multiple samples is tightly captured by the information-theoretic indistinguishability of $\widetilde{X}_0, \widetilde{X}_1$ under multiple samples. In particular, $k = \Theta\left (1/d^2_{\mathrm{H}}(\widetilde{X}_0, \widetilde{X}_1)\right )$ samples are both necessary and sufficient to distinguish $X_0, X_1$ with constant advantage. We can abstract this latter corollary as follows: 

\begin{definition}\label{def:pseudohellinger}
    For random variables $X_0, X_1$ over $\mathcal{X}$, $s \in \N$, $\eps > 0$, the \emph{$(s, \eps)$ pseudo-Hellinger distance} between $X_0$ and $X_1$ is the smallest $\delta$ such that there exist random variables $\widetilde{X}_0, \widetilde{X}_1$ such that:
    \begin{enumerate}
        \item $X_0$ is $\eps$-indistinguishable from $\widetilde{X}_0$ for circuits of size $s$.
        \item $X_1$ is $\eps$-indistinguishable from $\widetilde{X}_1$ for circuits of size $s$.
        \item $d_{\mathrm{H}}(\widetilde{X}_0, \widetilde{X}_1) \leq \delta$.
    \end{enumerate}
\end{definition}

With this definition in hand, we derive the following characterization of the indistinguishability of $X_0, X_1$, which is a computational analogue of Inequality~(\ref{ineq:hellinger}):

\begin{theorem}\label{thm:pseudohellinger}
    If $X_0, X_1$ have $(s, \eps)$ pseudo-Hellinger distance $\delta$, then for every $k$:
    \begin{enumerate}
        \item  $X_0^{\otimes k}, X_1^{\otimes k}$ are $\sqrt{2k \delta^2} + 2k \eps$-indistinguishable for circuits of size $s$.
        \item $X_0^{\otimes k}, X_1^{\otimes k}$ are $(1 - e^{-k\delta^2} - 2k\eps)$ distinguishable by circuits of size $s' = O(sk / \eps^6) + \poly(k / \eps)$.
    \end{enumerate}
\end{theorem}

Thus, the number of samples needed to distinguish $X_0, X_1$ with constant advantage is $\Theta(1 / \delta^2)$, where $\delta$ is the pseudo-Hellinger distance. As an immediate consequence, just like the traditional notion of Hellinger distance in the statistical distinguishability setting, our notion of pseudo-Hellinger distance gives an \emph{instance-optimal} characterization of the computational distinguishability of any pairs of random variables. This is a key benefit of our work over prior works \cite{HR08, geier2022tight}.

To prove Part (1) of \cref{thm:pseudohellinger}, we first observe that, by parts (1) and (2) of \cref{def:pseudohellinger} and a simple hybrid argument, the computational indistinguishability of $X_0^{\otimes k}$ and $X_1^{\otimes k}$ is bounded above by the computational indistinguishability of $\widetilde{X}_0^{\otimes k}$ and $\widetilde{X}_1^{\otimes k}$, plus $2 k \eps$. Next, the computational indistinguishability of $\widetilde{X}_0^{\otimes k}$ and $\widetilde{X}_1^{\otimes k}$ is upper-bounded by the total variation distance between $\widetilde{X}_0^{\otimes k}, \widetilde{X}_1^{\otimes k}$, which by Inequality~(\ref{ineq:hellinger}) is bounded above by $\sqrt{2 k d_H^2 (\widetilde{X}_0, \widetilde{X}_1)} = \sqrt{2 k \delta^2}$. This gives us the bound in Part (1) of the theorem.

To prove Part (2) of \cref{thm:pseudohellinger}, we 
use \cref{thm:main} to obtain $\widetilde{X}_0, \widetilde{X}_1$ that are $\eps$-indistinguishable from $X_0,X_1$ such that the computational distinguishability of $X_0^{\otimes k}, X_1^{\otimes k}$ by circuits of size $s'$ is at least $d_{\mathrm{TV}}(\widetilde{X}_0^{\otimes k}, \widetilde{X}_1^{\otimes k}) - 2 k \eps$. In turn, Inequality~(\ref{ineq:hellinger}) tells us that this is at least $1 - e^{-k \cdot d^2_{\mathrm{H}}(\widetilde{X}_0, \widetilde{X}_1)} - 2 k \eps$, which is then at least $ 1 - e^{-k \delta^2} -2k \eps$, where $\delta$ is the pseudo-Hellinger distance, $d_{\mathrm{H}}(\widetilde{X}_0, \widetilde{X}_1)$ over all $\widetilde{X}_0, \widetilde{X}_1$ that are $\eps$-indistinguishable from $X_0, X_1$.

Additionally, using Part 4 of \Cref{thm:main}, we can prove that there exists a $\widetilde{X}_0$ such that $k = \Theta\left (1/d^2_{\mathrm{H}}(\widetilde{X}_0, X_1)\right )$ samples are both necessary and sufficient to distinguish $X_0, X_1$ with constant 
advantage. When $X_1$ is uniform on $\mathcal{X}$, squared Hellinger distance $d^2_{\mathrm{H}}(\widetilde{X}_0, X_1)$ can be related to the R\'enyi $\frac{1}{2}$-entropy of $\widetilde{X}_0$, defined as follows: 
$$\mathrm{H}_{1/2}\left(\widetilde{X}_0\right) = 2 \log_2 \left( \sum_{i \in \mathcal{X}} \sqrt{\Pr[\widetilde{X}_0 = i]}\right).$$ 
Specifically, we have 
$$d_H^2\left(\widetilde{X}_0, X_1\right) = 1 - 2^{-\frac{1}{2} \left( \log |\mathcal{X}| - \mathrm{H}_{1/2}\left(\widetilde{X}_0\right)\right)} = \Theta\left( \min\left\{ \log|\mathcal{X}| - \mathrm{H}_{1/2}\left(\widetilde{X}_0\right), 1\right\}\right).$$
This allows us to create another suitable abstraction for the case where $X_1$ is uniform, beginning with the following definition of \textit{pseudo-R\'enyi entropy}.

\begin{definition}\label{def:pseudorenyi}
    For random variable $X_0$ over $\mathcal{X}$, $s \in \N$, $\eps > 0$, the \emph{$(s, \eps)$ pseudo-R\'enyi $\frac{1}{2}$-entropy} of $X_0$ is the smallest $r$ such that there exists a random variable $\widetilde{X}_0$ such that:
    \begin{enumerate}
        \item $X_0$ is $\eps$-indistinguishable from $\widetilde{X}_0$ for circuits of size $s$.
        \item $H_{1/2}\left(\widetilde{X}_0\right) = r$.
    \end{enumerate}
\end{definition}

This definition yields the following characterization of the indistinguishability of $X_0$ from the uniform distribution:

\begin{theorem}\label{thm:pseudorenyi}
    If $X_0$ is a distribution over $\mathcal{X}$ that has $(s, \eps)$ pseudo-R\'enyi $\frac{1}{2}$-entropy $r = \log|\mathcal{X}| - \delta$ and $\mathcal{U}$ is the uniform distribution over $\mathcal{X}$, then for every $k$:
    \begin{enumerate}
        \item  $X_0^{\otimes k}, \mathcal{U}^{\otimes k}$ are $O(\sqrt{k \delta}) + 2k \eps$-indistinguishable for circuits of size $s$.
        \item $X_0^{\otimes k}, \mathcal{U}^{\otimes k}$ are $(1 - e^{-\Omega(k \min\{\delta, 1\})} - 2k\eps)$ distinguishable by circuits of size $s' = O(sk / \eps^{12}) + \poly(k / \eps)$.
    \end{enumerate}
\end{theorem}

The number of samples needed to distinguish $X_0$ from uniform is $\Theta\left(\frac{1}{\delta} \right)$ where $\delta$ is the gap between the pseudo-R\'enyi $\frac{1}{2}$-entropy and its maximum possibly value (namely $\log|\mathcal{X}|$).

As mentioned, we also can deduce Geier's result \cite{geier2022tight} as stated above directly from \cref{thm:main}. A formal statement and comparison can be found in Theorem \ref{thm:geier-implication} and Section \ref{section:geier}, but we note that Geier's quantitative bound (the specific polynomial loss in circuit complexity) is better than ours.  In addition, Geier proves a version of the result for uniform distinguishers, whereas we only work with nonuniform distinguishers (i.e., boolean circuits).  Both of these limitations come from the currently known theorems about {\em multicalibration}, which is the main technical tool we use (as discussed below), and it is interesting problem to obtain multicalibration theorems that provide tight quantitative bounds and yield uniform-complexity results in applications such as ours. The benefit of using multicalibration is that it provides a direct translation between the computational and information-theoretic setting (e.g. as captured in \cref{thm:main}), which not only yields existing results as corollaries but also offers new ones, such as \cref{thm:pseudohellinger}.

\subsection{Asymptotic Complexity Formulations}\label{sec:asymptoticComplexity}

In the foundations of cryptography, it is common to state computational indistinguishability results in terms of asymptotic polynomial complexity.
In this section, we demonstrate the extensibility of our results to the asymptotic setting by presenting \cref{thm:pseudohellinger} in such a way, along with concrete definitions of this asymptotic regime. We start by formalizing indistinguishability in the asymptotic setting (i.e. with respect to ensembles of random variables):

\begin{definition}\label{def:computational-indistinguishability-ensembles}
    Let $X = \{ X_n \}_{n \in \N}$ and $Y = \{ Y_n \}_{n \in \N}$ be ensembles of random variables, where each $X_n, Y_n$ are supported on $\zo^{m(n)}$ and $m(n) = n^{O(1)}$. For $\eta: \N \rightarrow [0,1]$, we say that $X \equiv_{\eta}^{\mathrm{comp}} Y$ if for all $c$, there exists an $n_0$ such that for all $n \geq n_0$, $X_n$ is 
    $(\eta(n) + n^{-c})$-indistinguishable from $Y_n$ by circuits of size $n^c$. 

    Equivalently, we say $X \equiv_{\eta}^{\mathrm{comp}} Y$ if there exists $s(n) = n^{\omega(1)}$, $\eps(n) = n^{-\omega(1)}$ such that for all $n$ 
    $X_n$ is 
    $(\eta(n) + \eps(n))$-indistinguishable from $Y_n$ with respect to size $s(n)$ circuits.\footnote{The equivalence of these two formulations goes back to Bellare \cite{Bel02}.}

For simplicity, we will also use $X \equiv^{\mathrm{comp}} Y$ to denote that \emph{there exists} some function $\eta: \N \rightarrow [0,1]$, $\eta(n) = n^{-\omega(1)}$ such that $X \equiv^{\mathrm{comp}}_{\eta} Y$.
\end{definition}

We also generalize the notion of pseudo-Hellinger distance (\Cref{def:pseudohellinger}) to the setting of ensembles of random variables:

\begin{definition}
    For ensembles of random variables $X$ and $Y$, we say that $X$ and $Y$ have \textit{pseudo-Hellinger distance at most} $\delta$ (denoted $X \equiv_{\delta}^{cH} Y$) for a function $\delta: \N \rightarrow [0,1]$, if there exist ensembles $\widetilde{X} = \{\widetilde{X}_n \}, \widetilde{Y} = \{\widetilde{Y}_n\}$ such that 
    $X \equiv^{\mathrm{comp}} \widetilde{X}$, $Y \equiv^{\mathrm{comp}} \widetilde{Y}$, and $\forall n$, $d_H(\widetilde{X}_n, \widetilde{Y}_n) \leq \delta(n)$.
\end{definition}

Finally, we require a notion of the sample complexity required to distinguish ensembles:

\begin{definition}
    We say that ensembles $X, Y$ have \textit{computational sample complexity at least} $k: \N \rightarrow \N$ for $k(n) = n^{O(1)}$ if $X^k \equiv^{\mathrm{comp}}_{1/2} Y^k$ (where $X^k = \{X_n^{k(n)}\}$)\footnote{This choice of $1/2$ is arbitrary, and can be amplified through repetition.}. We denote this as $X \equiv_k^{cS} Y$.
\end{definition}

With these definitions, we can now state a corollary of our characterization of the indistinguishability of random variables in terms of their pseudo-Hellinger distance.

\begin{corollary}[Asymptotic Formulation of \cref{thm:pseudohellinger}]\label{cor:asymptotic}
    Let $X, Y$ be ensembles of random variables, let $\delta: \N \rightarrow [0,1]$, let $k: \N \rightarrow \N$ be a function such that $k(n) = n^{O(1)}$, and let $\delta(n) = n^{-O(1)}$. 
    
    Then:
    \begin{enumerate}
        \item If $X \equiv_{\delta}^{cH} Y$, then $X \equiv_k^{cS} Y$ for some $k(n) = \Omega \left ( \frac{1}{\delta(n)^2}\right ) $.
        \item If $X \equiv_k^{cS} Y$, then $X \equiv^{cH}_{\delta} Y$ for some $\delta(n) = O \left ( \frac{1}{\sqrt{k(n)}}\right )$.
    \end{enumerate}
\end{corollary}

We delegate the proof of this to the appendix.

\subsection{Technical Overview}\label{sec:overview}

Our work builds on the recent work of Casacuberta, Dwork, and Vadhan \cite{casacuberta2024complexity}. Driving inspiration from \cite{dwork2023pseudorandomness}, they showed how the recent notion of \textit{multicalibration} from the algorithmic fairness literature \cite{HKRR18} leads to simpler and more illuminating proofs of a number of fundamental known results in complexity theory and cryptography, such as the Impagliazzo Hardcore Lemma \cite{impagliazzo1995hard}, the Dense Model Theorem \cite{reingold2008dense}, and characterizations of pseudoentropy \cite{vadhan2012characterizing}. We show how multicalibration can be used to derive \textit{new} results about computational indistinguishability (Theorems \ref{thm:main} and \ref{thm:pseudohellinger}) and in general reduce computational reasoning to information-theoretic reasoning. Specifically, applying the Multicalibration Theorem \cite{HKRR18} to distinguishing problems in a way inspired by \cite{casacuberta2024complexity}, we obtain the following. Let $\widetilde{X}_{b|p(\widetilde{X}_b)=y}$ be the distribution where $x$ is sampled according to $\widetilde{X}_b$ conditioned on $p(x) = y$, for some function $p: \mathcal{X} \to [m]$. Let $p(X_0)$ be the distribution over $[m]$ where for $y \in [m]$, the probability that $p(X_0) = y$ is exactly $\sum_{x \in p^{-1}(y)} X_0(x)$.

\begin{theorem}\label{thm:MCproperties}
    For every pair of random variables $X_0, X_1$, every positive integer $s$, and every $\eps>0$, there exist random variables $\widetilde{X}_0, \widetilde{X}_1$ and a function $p : \mathcal{X} \to [m]$ for $m=O(1/\eps)$ such that:
    \begin{enumerate}[label=(\alph*)]
        \item $\widetilde{X}_0$ is $\eps$-indistinguishable from $X_0$ and $\widetilde{X}_1$ is $\eps$-indistinguishable from $X_1$ for circuits of size $s$.
        \item $p(X_0)$ is identically distributed to $p(\widetilde{X}_0)$ and $p(X_1)$ is identically distributed to $p(\widetilde{X}_1)$.
        \item For every $y \in [m]$, $\widetilde{X}_{0|p(\widetilde{X}_0)=y}$ is identically distributed to $\widetilde{X}_{1|p(\widetilde{X}_1)=y}$.
        \item $p$ is computable by circuits of size $O(s / \eps^6)$.
    \end{enumerate}
\end{theorem} 

Because the above theorem shows that $\widetilde{X}_0, X_0$ and $\widetilde{X}_1, X_1$ are $\eps$-indistinguishable for circuits of size $s$, then by a simple hybrid argument, $\widetilde{X}_0^{\otimes k}, X_0^{\otimes k}$ and $\widetilde{X}_1^{\otimes k}, X_1^{\otimes k}$ are $\eps k$-indistinguishable for circuits of size $s$. Thus, one direction of \cref{thm:main} clearly follows given \cref{thm:MCproperties}: size $s$ circuits cannot distinguish $X_0^{\otimes k}, X_1^{\otimes k}$ better than $d_{\mathrm{TV}}(\widetilde{X}_0^{\otimes k}, \widetilde{X}_1^{\otimes k}) + 2k\eps$.

However, the other direction is more subtle; namely, showing that one \emph{can} distinguish $X_0^{\otimes k}, X_1^{\otimes k}$ with advantage approaching $d_{\mathrm{TV}}(\widetilde{X}_0^{\otimes k}, \widetilde{X}_1^{\otimes k})$ with small circuits. For this, we heavily rely on items (b), (c), and (d)  from \cref{thm:MCproperties}, as well as the property that $m = O( 1/ \eps)$. The intuition is the following: consider an element $x$ that is sampled from either $\widetilde{X}_0, \widetilde{X}_1$, and let $p(x) = y$. By item (c), we know that once we condition on a value $p(x) = y$, $\widetilde{X}_0, \widetilde{X}_1$ are identically distributed. Thus, there is no information to be gained from the sample $x$ \emph{besides} the label of the partition that $x$ is in (i.e., the value $p(x)$).  In fact, this gives us a simple recipe for the \emph{optimal} distinguisher between $\widetilde{X}_0, \widetilde{X}_1$: for each sample $x$ we see, let us compute the value $p(x) = y$, and see how much more likely it is to sample an element in $p^{-1}(y)$ under $\widetilde{X}_0$ vs. $\widetilde{X}_1$. We can simply keep a running product over all of the samples $x_1, \dots , x_k$ of the form
\[
\prod_{j =1}^k \frac{\mathbb{P}_{x \sim \widetilde{X}_1}[p(x) = p(x_j)]}{\mathbb{P}_{x \sim \widetilde{X}_0}[p(x) = p(x_j)]}.
\]

If the above product is larger than $1$, this means that a sequence of values is more likely under $\widetilde{X}_1$, and if it is less than $1$, then this means a sequence of partitions of more likely under $\widetilde{X}_0$. In particular, calculating this simple expression and checking whether it is larger than $1$ is an \emph{optimal} distinguisher for $\widetilde{X}^{\otimes k}_0, \widetilde{X}^{\otimes k}_1$, as it is just the maximum likelihood decoder (that is to say, this decoder achieves distinguishing advantage $d_{\mathrm{TV}}(\widetilde{X}_0^{\otimes k}, \widetilde{X}_1^{\otimes k})$ between $\widetilde{X}^{\otimes k}_0, \widetilde{X}^{\otimes k}_1$). All that remains is to show that we can encode this distinguisher using small circuits. Here, we rely on point (d) of \cref{thm:MCproperties}: computing the function $p$ can be done with circuits of small size. Thus, for any element $x$, we compute the index of the partition that it is in $p(x)$, and feed this index into a look-up table (encoded non-uniformly) such that it returns the value $\frac{\mathbb{P}_{x \sim \widetilde{X}_1}[p(x) = p(x_j)]}{\mathbb{P}_{x \sim \widetilde{X}_0}[p(x) = p(x_j)]}$. Across all $k$ elements we see, we then must only keep track of the product, and return whether the value it computes is $\geq 1$. Here, we use the fact that $p$ has a small domain to prove that we can actually encode the look-up table mapping $y \in [m]$ to the values $\frac{\mathbb{P}_{x \sim \widetilde{X}_1}[p(x) = y]}{\mathbb{P}_{x \sim \widetilde{X}_0}[p(x) = y]}$ without using too large of a circuit. Of course, there are also issues with overflow and underflow in performing the numerical calculations, but we show this can be computed formally with small circuits in \cref{sec:efficient}. 

Finally, it remains to analyze the distinguisher we have constructed here for $\widetilde{X}^{\otimes k}_0, \widetilde{X}^{\otimes k}_1$ when we apply it to $X_0^{\otimes k}, X_1^{\otimes k}$. Here, we rely on Part (b) of \cref{thm:MCproperties}. Indeed, the distinguisher relies \emph{only} on the probabilities $p(\widetilde{X}_0), p(\widetilde{X}_1)$ (i.e., the probability mass placed on each value of $y$). \emph{But}, the distributions $X_0, X_1$ place exactly the same probability mass on each part of the partition (formally, $p(X_b) = p(\widetilde{X}_b)$). Thus, whatever distinguishing advantage our distinguisher achieves on $\widetilde{X}^{\otimes k}_0, \widetilde{X}^{\otimes k}_1$ is \emph{exactly the same} as its distinguishing advantage over $X_0^{\otimes k}, X_1^{\otimes k}$. This then yields the second part of \cref{thm:main}.

\subsubsection{Multicalibrated Partitions}

We now review the concept of multicalibration and describe how we prove \Cref{thm:MCproperties} from the Multicalibration Theorem.

Multicalibration is a notion that first arose in the algorithmic fairness literature. Roughly speaking, its goal is, given a predictor $h$ of an unknown function $g$, and a domain $\mathcal{X}$ (which is often thought of as a population of individuals), to ensure that every desired \emph{subpopulation} $S \subseteq \mathcal{X}$ receives calibrated predictions from $h$. The subpopulations are specified by their characteristic functions $f$, which came from a family $\mathcal{F}$. More formally, we say that for domain $\mathcal{X}$ and distribution $\mathcal{D}$ over $\mathcal{X}$, a function $g: \mathcal{X} \rightarrow [0,1]$, and a class $\mathcal{F}$ of functions $f: \mathcal{X} \rightarrow [0,1]$, $h$ is a \textit{$(\mathcal{F}, \eps)$ multicalibrated predictor} for $g$ with respect to $\mathcal{D}$, if $\forall f \in \mathcal{F}$, and $\forall v \in \mathrm{image}(h)$:
\[
\left | \E_{x \sim \mathcal{D}}[f(x) \cdot (g(x) - h(x)) | h(x) = v] \right | \leq \eps.
\]

The sets $\{h^{-1}(v) : v \in \mathrm{Image}(h)\}$ define a partition of $\mathcal{X}$, which gives rise to the following more convenient formulation.  Let $\mathcal{D}|_{P}$ denote the distribution $\mathcal{D}$ conditioned on being in the subset $P \subseteq \mathcal{X}$ of the domain.

We use a partition-based characterization of multicalibration as used in previous works \cite{GKSZ22, GKRSW22, dwork2023pseudorandomness, casacuberta2024complexity}:
\begin{definition}[Multicalibration \cite{HKRR18} as formulated in \cite{casacuberta2024complexity}]
    Let $g: \mathcal{X} \rightarrow [0,1]$ be an arbitrary function, $\mathcal{D}$ be a probability distribution over $\mathcal{X}$, and for an integer $s \in \mathbb{Z}^+$, let $\mathcal{F}^{(s)}$ be the class of functions $f: \mathcal{X} \rightarrow [0,1]$ computable by size $s$ circuits. Let $\eps, \gamma > 0$ be constants. Then, we say that a partition $\mathcal{P}$ of $\mathcal{X}$ is \textit{$(\mathcal{F}^{(s)}, \eps, \gamma)$-approximately multicalibrated for $g$ on $\mathcal{D}$} if for every $f \in \mathcal{F}^{(s)}$ and every $P \in \mathcal{P}$ such that $ \mathbb{P}_{x \sim \mathcal{D}}[x \in P] \geq \gamma$ it holds that 
    \[
    \left| \E_{x \sim \mathcal{D}|_{P}} [ f(x) \cdot (g(x) - v_P)]\right| \leq \eps,
    \]
    where we define $v_P := \E_{x \sim \mathcal{D}|_{P}}[g(x)]$.
\end{definition}

For the class $\mathcal{F}^{(s)}$ of functions computable by size $s$ circuits, let $\mathcal{F}^{(s,k)}$ denote the set of partitions $\mathcal{P}$ such that there exists $\hat{f} \in \mathcal{F}^{(s)}$, $\hat{f}: \mathcal{X} \rightarrow [k]$ and $\mathcal{P} = \{\hat{f}^{-1}(1), \dots ,  \hat{f}^{-1}(k)\}$.

The result of H{\'{e}}bert-Johnson, Kim, Reingold, and Rothblum \cite{HKRR18} can be stated as follows in the language of partitions: 

\begin{theorem}[Multicalibration Theorem \cite{HKRR18}]
    Let $\mathcal{X}$ be a finite domain, for an integer $s \in \mathbb{Z}^+$, let $\mathcal{F}^{(s)}$ be the class of functions $f: \mathcal{X} \rightarrow \{0,1\}$ computable by size $s$ circuits, let $g:\mathcal{X} \rightarrow [0,1] $ be an arbitrary function, $\mathcal{D}$ be a probability distribution over $\mathcal{X}$, and $\eps, \gamma > 0$. Then, there exists a $(\mathcal{F}_s, \eps, \gamma)$-approximately multicalibrated partition $\mathcal{P}$ of $\mathcal{X}$ for $g$ on $\mathcal{D}$ such that $\mathcal{P} \in \mathcal{F}^{(W, k)}$, where $W = O(s / \eps^6) + O(1 / (\eps^{12} \gamma) \cdot \log(|\mathcal{X}|/\eps))$ and $k = O(1/\eps)$.
\end{theorem}

In particular, one way to understand the above definition is that the partition $\mathcal{P}$ breaks the domain into parts, such that \emph{within} each part, for all functions $f \in \mathcal{F}$, the function $g$ is $\eps$-indistinguishable from its expected value. 

Recently, this perspective on multicalibration has proven to be quite fruitful, with applications to machine learning \cite{GKRSW22}, graph theory \cite{dwork2023pseudorandomness}, and complexity theory and cryptography \cite{dwork2023pseudorandomness, casacuberta2024complexity}. Philosophically, the work of \cite{casacuberta2024complexity} showed how to reprove (and strengthen) theorems of average-case hardness and indistinguishability using multicalibration (specifically, the Impagliazzo Hardcore Lemma \cite{impagliazzo1995hard}, the Dense Model Theorem \cite{reingold2008dense}, and characterizations of pseudoentropy \cite{vadhan2012characterizing}). One consequence of their work was a ``loose'' intuition that multicalibration translates questions about computational indistinguishability into questions about statistical indistinguishability.
Our goal is to show this translation more clearly and generally, and underline directly how multicalibration reduces an understanding of computational indistinguishability to an understanding of statistical indistinguishability (as captured by \cref{thm:main}). 

\subsubsection{Invoking Multicalibration}

As mentioned above, multicalibration has already found a host of applications. However, for the specific purpose of distinguishing two distributions $X_0, X_1$, it is not immediately clear how to apply this framework. Of course, we can naturally view each distribution as given by its probability mass function from $\mathcal{X}$ to $[0,1]$, but in order to create multicalibrated partitions, we need a function $g$ with respect to which the partitions should be calibrated. For this, our first observation is to use the following function $g$:
\[
g(x) = \begin{cases}
    0 ~~ \text{with probability } \frac{\mathbb{P}[X_0 = x]}{\mathbb{P}[X_0 = x] + \mathbb{P}[X_1 = x]} \\
    1 ~~ \text{with probability } \frac{\mathbb{P}[X_1 = x]}{\mathbb{P}[X_0 = x] + \mathbb{P}[X_1 = x]}.
\end{cases}
\]

Roughly speaking, the function $g$ is motivated by looking at the distribution $\mathcal{D} = \frac{1}{2}(X_0 + X_1)$. In this distribution, the probability of seeing an element $x$ is $(\mathbb{P}[X_0 = x] + \mathbb{P}[X_1 = x]) / 2$. However, we can also understand the sampling procedure as first choosing one of $X_0, X_1$ with probability $1/2$, and then sampling from the chosen distribution. The probability that $x$ comes from $X_0$ is then $\mathbb{P}[X_0 = x] / 2$, and similarly for $X_1$. $\E_{x \sim \mathcal{D}}[g(x)]$ is thus measuring the relative fraction of the time that an element $x$ comes from $X_1$.

Importantly, $g$ is now a well-defined function with respect to which we can construct a multicalibrated partition. Then we will define $\widetilde{X}_b$ for $b \in \{0, 1\}$ by replacing $X_b|_P$ with $\mathcal{D}|_P$.

We define this procedure formally below:

\begin{definition}\label{def:tilde_variables}
    For a pair of random variables $X_0, X_1$,  
    positive integer $s$, and parameter $\eps$,
    we define the distribution $\mathcal{D}$, function $g$, random variables $\widetilde{X}_0, \widetilde{X}_1$ as follows.
    \begin{enumerate}[label=(\alph*)]
        \item Let the distribution $\mathcal{D}$ be as follows: To sample from $\mathcal{D}$, first pick $B \sim \{0, 1\}$ uniformly at random. Output a sample $x \sim X_B$. Note that $$\mathbb{P}_{x \sim \mathcal{D}}[x] = \frac{1}{2} \mathbb{P}[X_0 = x] + \frac{1}{2}\mathbb{P}[X_1 = x].$$

        For a subset of the domain $\mathcal{S} \subseteq \mathcal{X}$, let $\mathcal{D}_{| \mathcal{S}}$ be the conditional distribution of $\mathcal{D}$ over the set $\mathcal{S}$.
        \item We then define the randomized function $g$ as:
        $$g(x) = \begin{cases}
            0 ~~ \text{with probability } \frac{\mathbb{P}[X_0 = x]}{\mathbb{P}[X_0 = x] + \mathbb{P}[X_1 = x]} \\
            1 ~~ \text{with probability } \frac{\mathbb{P}[X_1 = x]}{\mathbb{P}[X_0 = x] + \mathbb{P}[X_1 = x]}.
        \end{cases}$$
        \item Random variables $\widetilde{X}_0, \widetilde{X}_1$: Consider the multicalibrated partition $\mathcal{P} = \{P_i\}$ guaranteed by the Multicalibration Theorem when applied to the function $g$, distribution $\mathcal{D}$ over domain $\mathcal{X}$, class of functions $f: \mathcal{X} \to [0, 1]$ computable by size $s$ circuits, and parameters $\epsilon$ and $\gamma = \epsilon^2$. Given the multicalibrated partition, construct the random variables $\widetilde{X}_b$ as follows: to sample according to $\widetilde{X}_b$, first choose a piece of the partition $P_i \in \mathcal{P}$, where $P_i$ is chosen with probability $\mathbb{P}[X_b \in P_i]$. Then sample $x \sim \mathcal{D}|_{P_i}$.
    \end{enumerate}
\end{definition}

With this definition, we can provide an informal overview of the proof of \cref{thm:MCproperties}. 
We start with item (b), as its proof is essentially definitional. Recall that this item states that $p(X_0)$ and $p(\widetilde{X}_0)$ are identically distributed (and analogously for $X_1$). This follows because in the construction of the random variable $\widetilde{X}_0$, we sample an element from part $P_i$ with probability $\Pr[X_0 \in P_i]$, and thus necessarily, the probability mass placed on each part $P_i$ is identical between the two distributions. Next, we can also see that part (c) is definitional. In the construction of $\widetilde{X}_0$ and $\widetilde{X}_1$, conditioned on being in the piece $P_i$, each of the marginal distributions is exactly $\mathcal{D}_{P_i}$, and thus the marginal distributions match. 
Next, to conclude item (d) of \cref{thm:MCproperties}, recall that the partition function $p$ is exactly returned by the Multicalibration Theorem. Here, we can take advantage of the fact that multicalibrated partitions are efficiently computable given oracle access to the underlying set of test functions (in this case, size $\leq s$ circuits). By replacing these oracles with the size $\leq s$ circuits, this yields an overall complexity of $O(s / \eps^2)$ for the circuit size required to compute $p$. 
Finally, it remains to prove item (a) of \cref{thm:MCproperties}. Because the partition $\mathcal{P}$ that is returned is multicalibrated with respect to our function $g$, by applying Theorem 2.15 of \cite{casacuberta2024complexity} and following the approach of the DMT$++$ proof of \cite{casacuberta2024complexity}, we can show that for every part $P_i \in \mathcal{P}$ and every circuit of size $\leq s$, the distributions $X_{0}|_{P_i}$ and $\widetilde{X}_{0}|_{P_i}$ are close to indistinguishable (and likewise for $X_1$). By combining this ``local'' indistinguishability across all parts of the partition, this then yields the overall indistinguishability between $\widetilde{X}_0, \widetilde{X}_1$.

It is worth comparing our use of the Multicalibration Theorem to the use of Impagliazzo's Hardcore Theorem~\cite{impagliazzo1995hard,hol05} in some past works on computational indistinguishability and pseudorandomness \cite{HR08, MT10}. Applied to the same function $g$ above, the Hardcore Theorem says that if $X_0, X_1$ are $(s', 1-\delta)$
-indistinguishable, for $s' = O(s\cdot \poly(1/\eps,1/\delta))$, then there is a subset $H \subseteq X$ such that $\Pr_{x \in X_0}[x \in H] =\Pr_{x \in X_1}[x \in H]=\delta$ such that $X_0 |_H$ and $X_1|_H$ are $(s, \eps)$-indistinguishable. Replacing $X_0 |_H$ and $X_1|_H$ with their average as above, we obtain $\widetilde{X}_0, \widetilde{X}_1$ that are $(s, \eps)$-indistinguishable from $X_0, X_1$ respectively, and such that $d_{\mathrm{TV}}(\widetilde{X}_0, \widetilde{X}_1) \leq 1-\delta.$ Thus, the Hardcore Lemma tightly captures the single-sample computational indistinguishability of $X_0$ and $X_1$ by the single-sample information-theoretic indistinguishability of $\widetilde{X}_0$ and $\widetilde{X}_1$. But it does not seem to capture enough information to obtain an instance-optimal characterization of the multiple-sample indistinguishability, as we do. Concretely, the Hardcore Lemma requires assuming some initial average-case hardness of $g$ (which comes from the weak indistinguishability of $X_0$ and $X_1$) and only gives us indistinguishability from an information-theoretic counterpart on a small subset $H \subseteq \mathcal{X}$, which is not guaranteed to be efficiently recognizable. In contrast, the partition given by the Multicalibration Theorem covers the entire domain $\mathcal{X}$ with efficiently recognizable sets, and this is crucial for our instance-optimal characterization.

\subsection{Organization of the paper}

Sections \ref{section:apply-MC}, \ref{sec:mainLB},  and \ref{section:one-intermediate} prove the different components of \Cref{thm:main}. In \Cref{section:apply-MC}, we apply the Multicalibration Theorem to prove \Cref{thm:MCproperties}. \Cref{thm:MCproperties} Part (a) is equivalent to \Cref{thm:main} Part (1). \Cref{sec:mainLB} proves \Cref{thm:main} Parts (2) and (3); see \Cref{thm:mainRestated} for a more general version of the statement we prove, which encompasses indistinguishability against families of functions beyond size $s$ circuits. \Cref{section:one-intermediate} proves \Cref{thm:main} Part (4); see \Cref{thm:computational-indistinguishability-3} for the general version of the statement, again for broader families of functions. These three sections all rely on \Cref{thm:MCproperties}.

\Cref{section:geier} shows how our results and analysis imply a result comparable to the main theorem proven in \cite{HR08, geier2022tight}. \Cref{sec:pseudo-hellinger-renyi} proves \Cref{thm:pseudohellinger} and \Cref{thm:pseudorenyi}, which characterize distinguishability in terms of pseudo-Hellinger distance and pseudo-R\'enyi $\frac{1}{2}$-entropy (as defined in \Cref{def:pseudohellinger} and \Cref{def:pseudorenyi}). Lastly, \Cref{section:general-product-distributions} generalizes our results to the distinguishability of general product distributions (instead of $X_0^{\otimes k}$ versus $X_1^{\otimes k}$).

\section{Preliminaries}

\subsection{Distinguishability}

Hellinger distance can be used to characterize the ability to distinguish two distributions when we receive multiple samples:

\begin{claim}\label{claim:hellinger_facts}[See \cite{ste1983, Woo2019}]
\begin{enumerate}
    \item $d_H^2(X, Y) \leq d_{\mathrm{TV}}(X, Y) \leq \sqrt{2} \cdot d_H(X, Y)$.
    \item $d_H^2(X^{\otimes m}, Y^{\otimes m}) = 1 - (1 - d_H^2(X, Y))^m \in [1 -e^{-m \cdot d_H^2(X, Y)}, m \cdot d_H^2(X, Y)]$.
\end{enumerate}
\end{claim}
Combining the above gives us that 
\begin{equation}\label{equation:TVD-Hellinger-inequalities}
    1 - e^{-m \cdot d_H^2(X, Y)} \leq d_{\mathrm{TV}}(X^{\otimes m}, Y^{\otimes m}) \leq \sqrt{2m} \cdot d_H(X, Y).
\end{equation}

As a corollary, this implies that $m = \Theta(1/d_H^2(X, Y))$ samples are necessary and sufficient to increase the total variation distance $d_{\mathrm{TV}}(X^{\otimes m}, Y^{\otimes m})$ up to a constant.

\begin{definition}\label{renyi}
    Let $X$ be a random variable over $\mathcal{X}$. The {\em R\'enyi entropy of order $1/2$} of $X$ is defined as:
    $$\mathrm{H}_{1/2}(X) = 2 \log_2 \left( \sum_{i \in \mathcal{X}} \sqrt{\Pr[X = i]}\right).$$
\end{definition}

It can be shown that $\mathrm{H}_{1/2}(X) \leq \log_2|\mathcal{X}|$ with equality if and only if $X$ is uniform on $\mathcal{X}$. More generally, the gap in this inequality exactly measures the Hellinger distance to the uniform distribution on $\mathcal{X}$.

\begin{claim}\label{renyi-and-hellinger}
    Let $X$ be a random variable over $\mathcal{X}$ and let $\mathcal{U}$ be the uniform distribution over $\mathcal{X}$. Then 
    $$d_H^2\left(X, \mathcal{U}\right) = 1 - \sqrt{\frac{2^{\mathrm{H}_{1/2}\left(X\right)}}{|\mathcal{X}|}}.$$
\end{claim}

\begin{proof} We can rewrite $d_H^2\left(X, \mathcal{U}\right)$ as follows:
    $$d_H^2\left(X, \mathcal{U}\right) = \frac{1}{2} \sum_{i \in \mathcal{X}} \left( \sqrt{\mathbb{P}[X = i]} - \frac{1}{\sqrt{|\mathcal{X}|}}\right)^2  = \frac{1}{2} \left( 1 + 1 - 2 \sqrt{\frac{\mathbb{P}[X = i]}{|\mathcal{X}|}}\right)$$
    $$= 1 - \frac{1}{\sqrt{|\mathcal{X}|}} \sum_{i \in \mathcal{X}} \sqrt{\mathbb{P}[X = i]} = 1 - \sqrt{\frac{2^{\mathrm{H}_{1/2}\left(X\right)}}{|\mathcal{X}|}}. \eqno \qedhere$$
\end{proof}

\begin{definition}[Statistical indistinguishability]
     Random variables $X$ and $Y$ are called $\eps$-\textit{statistically indistinguishable} if $d_{\mathrm{TV}}(X, Y) \leq \eps$.
\end{definition}

We define computational indistinguishability for a class of functions $\mathcal{F}$:

\begin{definition}[Computational indistinguishability]
    Random variables $X$ and $Y$ over domain $\mathcal{X}$ are $\eps$-\textit{indistinguishable} with respect to $\mathcal{F}$ if for every $f \in \mathcal{F}$:
    $$\left|\sum_{x \in \mathcal{X}} f(x) \cdot \left(\mathbb{P}[X = x] - \mathbb{P}[Y = x]\right) \right| \leq \eps.$$
    Random variables $X$ and $Y$ are $\eps$-\textit{distinguishable} with respect to $\mathcal{F}$ if for every $f \in \mathcal{F}$:
    $$\left|\sum_{x \in \mathcal{X}} f(x) \cdot \left(\mathbb{P}[X = x] - \mathbb{P}[Y = x]\right) \right| \geq \eps.$$
\end{definition}

\begin{definition}[Oracle-aided circuits and partitions]
For a class $\mathcal{F}$ of functions, let $\mathcal{F}_{q, t}$ be the class of functions that can be computed by an oracle-aided circuit of size $t$ with $q$ oracle gates, where each oracle gate is instantiated with a function from $\mathcal{F}$. Let $\mathcal{F}_{q, t, k}$ denote the set of partitions $\mathcal{P}$ such that there exists $\hat{f} \in \mathcal{F}_{q, t}$, $\hat{f}: \mathcal{X} \rightarrow [k]$ and $\mathcal{P} = \{\hat{f}^{-1}(1), \dots ,  \hat{f}^{-1}(k)\}$.
\end{definition}

Note that when we restrict our attention to $\mathcal{F}$ being the set of functions computable by size $s$ circuits, then $\mathcal{F}_{q, t}$ is simply the set of functions computable by size $t + q \cdot s$ circuits. The discussion in the introduction is using exactly this simplification. 

\subsection{Multicalibration}

We now recall the definition of multicalibration \cite{HKRR18, GKRSW22, GKSZ22, CDV23}, but state it in its full generality (with respect to arbitrary classes of functions $\mathcal{F}$):

\begin{definition}[Multicalibration \cite{HKRR18} as formulated in \cite{casacuberta2024complexity}]\label{def:multicalibration}
    Let $g: \mathcal{X} \rightarrow [0,1]$ be an arbitrary function, $\mathcal{D}$ be a probability distribution over $\mathcal{X}$, and $\mathcal{F}$ be a class of functions $f \in \mathcal{F}, f: \mathcal{X} \rightarrow [0,1]$. Let $\eps, \gamma > 0$ be constants. A partition $\mathcal{P}$ of $\mathcal{X}$ is $(\mathcal{F}, \eps, \gamma)$-\textit{approximately multicalibrated for $g$ on $\mathcal{D}$} if for every $f \in \mathcal{F}$ and every $P \in \mathcal{P}$ satisfying $ \mathbb{P}_{x \sim \mathcal{D}}[x \in P] \geq \gamma$ it holds that 
    \[
    \left| \E_{x \sim \mathcal{D}|_{P}} [ f(x) \cdot (g(x) - v_P)]\right| \leq \eps,
    \]
    where we define $v_P := \E_{x \sim \mathcal{D}|_{P}}[g(x)]$.
\end{definition}

As mentioned above, \cite{HKRR18} shows how to construct multicalibrated partitions with low-complexity relative to the class of functions $\mathcal{F}$. Their main result can be stated as:

\begin{theorem}[Multicalibration Theorem \cite{HKRR18}]\label{thm:multicalibrationGeneralFunctions}
    Let $\mathcal{X}$ be a finite domain, $\mathcal{F}$ a class of functions, with $f \in \mathcal{F}: f: \mathcal{X} \rightarrow [0,1]$, $g:\mathcal{X} \rightarrow [0,1] $ an arbitrary function, $\mathcal{D}$ a probability distribution over $\mathcal{X}$, and $\eps, \gamma > 0$. Then, there exists a $(\mathcal{F}, \eps, \gamma)$-approximately multicalibrated partition $\mathcal{P}$ of $\mathcal{X}$ for $g$ on $\mathcal{D}$ such that $\mathcal{P} \in \mathcal{F}_{q, t, k}$ for $t = O(1 / (\eps^{12} \gamma) \cdot \log(|\mathcal{X}|/\eps))$, $q = O(1 / \eps^6)$, and $k = O(1 / \eps)$.
\end{theorem}

\section{Applying the Multicalibration Theorem to Prove \Cref{thm:MCproperties}}\label{section:apply-MC}

In this section, we apply the Multicalibration Theorem to prove the following theorem, which is a generalization of \Cref{thm:MCproperties}. In later sections, we will use this theorem to prove the different components of \Cref{thm:main}.

\begin{theorem}[General version of \Cref{thm:MCproperties}]\label{thm:MCproperties_restatement}
    For every pair of random variables $X_0, X_1$, every family $\mathcal{F}$ of functions $f: \mathcal{X} \to [0, 1]$, and every $\eps>0$, there exist random variables $\widetilde{X}_0, \widetilde{X}_1$ and a function $p : \mathcal{X} \to [m]$ for $m=O(1/\eps)$ such that:
    \begin{enumerate}[label=(\alph*)]
        \item $\widetilde{X}_0$ is $\eps$-indistinguishable from $X_0$ and $\widetilde{X}_1$ is $\eps$-indistinguishable from $X_1$ for functions $f \in \mathcal{F}$.
        \item $p(X_0)$ is identically distributed to $p(\widetilde{X}_0)$ and $p(X_1)$ is identically distributed to $p(\widetilde{X}_1)$.
        \item For every $y \in [m]$, $\widetilde{X}_{0|p(\widetilde{X}_0)=y}$ is identically distributed to $\widetilde{X}_{1|p(\widetilde{X}_1)=y}$.
        \item $p$ is computable by functions in $\mathcal{F}_{O(1 / \eps^6), O(1 / (\eps^{12}) \cdot \log(|\mathcal{X}|)\cdot\log(|\mathcal{X}| / \eps) )}$. 
    \end{enumerate}
\end{theorem} 

\paragraph{Definitions and notation used throughout the proof.} The random variables $\widetilde{X}_0, \widetilde{X}_1$ and function $p: \mathcal{X} \to [m]$ in \cref{thm:MCproperties_restatement} are constructed as follows, by drawing a connection to the multicalibrated partition guaranteed by the Multicalibration Theorem.

\begin{definition}\label{def:tilde_variables_general}(General version of \cref{def:tilde_variables})
    For a pair of random variables $X_0, X_1$, class $\mathcal{F}$ of functions, and parameter $\eps$,
    we define the family of functions $\mathcal{F}'$, distribution $\mathcal{D}$, function $g$, random variables $\widetilde{X}_0, \widetilde{X}_1$, and function $p: \mathcal{X} \to [m]$ as follows.
    \begin{enumerate}[label=(\alph*)]
        \item Let $\mathcal{F}'$ be a family of functions such that $\mathcal{F}_{c , c\log |\mathcal{X}|} \subseteq \mathcal{F}'$. For example, if $\mathcal{F}$ corresponds to size $\leq s$ circuits, then $\mathcal{F}'$ is the family of functions given by circuits of size at most $s \cdot c + c \cdot \log |\mathcal{X}|$, for some universal constant $c$.
        \item Let the distribution $\mathcal{D}$ be as follows: To sample from $\mathcal{D}$, first pick $B \sim \{0, 1\}$ uniformly at random. Output a sample $x \sim X_B$. Note that $$\mathbb{P}_{x \sim \mathcal{D}}[x] = \frac{1}{2} \mathbb{P}[X_0 = x] + \frac{1}{2}\mathbb{P}[X_1 = x].$$

        For a subset of the domain $\mathcal{S} \subseteq \mathcal{X}$, let $\mathcal{D}|_{\mathcal{S}}$ be the conditional distribution of $\mathcal{D}$ over the set $\mathcal{S}$.
        \item We then define the randomized function $g$ as:
        $$g(x) = \begin{cases}
            0 ~~ \text{with probability } \frac{\mathbb{P}[X_0 = x]}{\mathbb{P}[X_0 = x] + \mathbb{P}[X_1 = x]} \\
            1 ~~ \text{with probability } \frac{\mathbb{P}[X_1 = x]}{\mathbb{P}[X_0 = x] + \mathbb{P}[X_1 = x]}.
        \end{cases}$$
        \item Random variables $\widetilde{X}_0, \widetilde{X}_1$: Consider the multicalibrated partition $\mathcal{P} = \{P_i\}$ guaranteed by the Multicalibration Theorem when applied to the function $g$, distribution $\mathcal{D}$ over domain $\mathcal{X}$, class $\mathcal{F}'$ of functions $f: \mathcal{X} \to [0, 1]$, and parameters $\epsilon$ and $\gamma = \epsilon^2$. Given the multicalibrated partition, construct the random variables $\widetilde{X}_b$ as follows: to sample according to $\widetilde{X}_b$, first choose a piece of the partition $P_i \in \mathcal{P}$, where $P_i$ is chosen with probability $\mathbb{P}[X_b \in P_i]$. Then sample $x \sim \mathcal{D}|_{P_i}$.

        Let $m = |\mathcal{P}|$ be the number of parts in this partition $\mathcal{P}$.
        \item We let $p: \mathcal{X} \to [m]$ be the function that returns which part of the multicalibrated partition $\mathcal{P}$ an element $x \in \mathcal{X}$ is in. That is, if $x \in P_i$ for $P_i \in \mathcal{P}$, $p(x) = i$.
    \end{enumerate}
\end{definition}

Given this setup of $\widetilde{X}_0$, $\widetilde{X}_1$, and $p$, we are now ready to prove the different parts of \Cref{thm:MCproperties_restatement}.

\subsection{Proof of Part (a)}

We first analyze the behavior of the random variables over parts of the partition given by the Multicalibration Theorem. We begin by studying the indistinguishability of $X_{0}|_{P_i}$ versus $X_{1}|_{P_i}$, relying on the following lemma from \cite{casacuberta2024complexity}.

\begin{lemma}[Lemma 2.15 \cite{casacuberta2024complexity}]\label{lemma:2.15_casacuberta}
    Let $\mathcal{H} = \{h : \mathcal{X} \to [0, 1]\}$ be a class of functions that is closed under negation and contains the all-zero ($h(x) = 0$ for all $x$) and all-one ($h(x) = 1$ for all $x$) functions. Let $\mathcal{D}$ be a distribution over $\mathcal{X}$ and consider $\eps > 0$. Let $\mathcal{H}'$ be any family of functions such that $\mathcal{H}'_{c \log |\mathcal{X}|, c} \subseteq \mathcal{H}$ for a universal constant $c$.
    
    Suppose that $g: \mathcal{X} \to [0, 1]$ is identified with a randomized Boolean-valued function and is $(\mathcal{H}, \eps)$-indistinguishable from the constant function $v := \mathbb{E}_{x \sim \mathcal{D}}[g(x)]$. Then the distribution $\mathcal{D}_{| g(x) = 1}$ is $\left(\mathcal{H}', \frac{\eps}{v (1 - v)}\right)$-indistinguishable from $\mathcal{D}_{| g(x) = 0}$ for $\mathcal{H}'$.
\end{lemma}

\begin{lemma}\label{lem:X0Pi_vs_X1Pi}
     For random variables $X_0, X_1$ and family of functions $\mathcal{F}$, consider the construction of the family of functions $\mathcal{F}'$, the distribution $\mathcal{D}$, function $g$, and partition $\mathcal{P}$ as in \cref{def:tilde_variables_general}. For $P_i \in \mathcal{P}$, let $v_{P_i} = \mathbb{E}_{x \sim \mathcal{D}|_{P_i}}[g(x)]$.
     
     For all $P_i \in \mathcal{P}$ such that $\mathbb{P}_{x \sim \mathcal{D}}[x \in P_i] \geq \gamma$, $X_{0}|_{P_i}$ is $(\mathcal{F}, \frac{\eps}{v_{P_i} (1-v_{P_i})})$-indistinguishable from $X_{1}|_{P_i}$.
\end{lemma}

\begin{proof}

    Similarly to the approach in the proof of Theorem 5.3 (DMT++) in \cite{casacuberta2024complexity}, we consider the distribution $\mathcal{D}$ that picks $B \sim \{0,1\}$ at random and outputs a sample of $x \sim X_B$, and the randomized function $g$ that outputs $B$. This precisely corresponds to the choice of $\mathcal{D}, g$ corresponding to $X_0, X_1$ and $\mathcal{F}'$ from Definition \ref{def:tilde_variables_general}. As in the proof of Theorem 5.3 in \cite{casacuberta2024complexity}, we will show that for all $P_i \in \mathcal{P}$, $\mathcal{D}_{P_i |  g(x) = 0} = X_{0}|_{P_i}$ and $\mathcal{D}_{P_i | g(x) = 1} = X_{1}|_{P_i}$; we can therefore use Lemma \ref{lemma:2.15_casacuberta} (setting $\mathcal{H} = \mathcal{F}'$) to transfer the indistinguishability of $g$ from $v_{P_i}$ over each part of the partition $P_i \in \mathcal{P}$ to the indistinguishability of $X_{0}|_{P_i}$ and $X_{1}|_{P_i}$. More specifically, the Multicalibration Theorem guarantees that $\mathcal{P}$ is a low-complexity partition with $O(1/\eps)$ parts such that, on each $P_i \in \mathcal{P}$ with $\mathbb{P}_{x \sim \mathcal{D}}[x \in P_i] \geq \gamma$, $g$ is $\eps$-indistinguishable from the corresponding constant function $v_{P_i} := \mathbb{E}_{x \sim \mathcal{D}_{P_i}}[g(x)]$. Applying Lemma \ref{lemma:2.15_casacuberta} with $\mathcal{H} $  set to be $\mathcal{F}'$ therefore implies that for each part $P_i$ of the partition $\mathcal{P}$ such that $\mathbb{P}_{x \sim \mathcal{D}}[x \in P_i] \geq \gamma$, $X_{1}|_{P_i}$ is $(\mathcal{F}, \frac{\eps}{v_{P_i} (1-v_{P_i})})$-indistinguishable from $X_{0}|_{P_i}$ for any class of functions $\mathcal{F}$ such that $\mathcal{F}_{c \log |\mathcal{X}|, c} \subseteq \mathcal{F}'$, for a universal constant $c$.

    We work out the details for showing $\mathcal{D}_{P_i |g(x) = 1} = X_{1}|_{P_i}$ below, and the proof of $\mathcal{D}_{P_i |g(x) = 0} = X_{0}|_{P_i}$ follows similarly. In what follows, let $\mathbb{P}_{\text{rand}(g)}$ denote that the probability of an event is taken over the randomness of the randomized Boolean-valued function $g$.
    
    We see that
\begin{equation}\label{equation:condition-on-g-1}
    \mathbb{P}_{x \sim \mathcal{D}_{P_i|g(x) = 1}}\left[x\right] = \frac{\mathbb{P}_{\text{rand}(g)}[g(x) = 1 | x] \cdot \mathbb{P}_{x \sim D_{P_i}}[x]}{\mathbb{P}_{\text{rand}(g), x \sim \mathcal{D}_{P_i}}[g(x) = 1]}.
\end{equation}
Now, the denominator equals:
$$\mathbb{P}_{\text{rand}(g), x \sim \mathcal{D}|_{P_i}}[g(x) = 1] = \frac{1}{2} \frac{1}{\mathbb{P}_{\mathcal{D}}[P_i]}\sum_{x \in \mathcal{X}} \mathbb{P}_{x \sim \mathcal{D}}[x] \cdot \frac{\mathbb{P}[X_1 = x]}{\mathbb{P}[X_0 = x] + \mathbb{P}[X_1 = x]} $$
$$
= \frac{\mathbb{P}[X_1 \in P_i]}{\mathbb{P}[X_0 \in P_i] + \mathbb{P}[X_1 \in P_i]}.$$
Similarly, the numerator equals:
$$\mathbb{P}_{\text{rand}(g)}[g(x) = 1 | x] \cdot \mathbb{P}_{x \sim \mathcal{D}|_{P_i}}[x] = \frac{\mathbb{P}[X_1 = x]}{\mathbb{P}[X_0 \in P_i] + \mathbb{P}[X_1 \in P_i]}.$$
Therefore, Equation (\ref{equation:condition-on-g-1}) gives:
$$\mathbb{P}_{x \sim \mathcal{D}_{P_i |g(x) = 1}}\left[x\right] = \frac{\mathbb{P}[X_1 = x]}{\mathbb{P}[X_1 \in P_i]} = \mathbb{P}[X_{1}|_{P_i} = x],$$
which means that $\mathcal{D}_{P_i |g(x) = 1}$ is equivalent to $X_{1}|_{P_i}$.
\end{proof}

We next prove that we can represent $\mathcal{D}_{P_i}$ as a convex combination of $X_{0}|_{P_i}$ and $X_{1}|_{P_i}$.

\begin{lemma}\label{lem:D-and-X0X1}
For random variables $X_0, X_1$ and family of functions $\mathcal{F}$, consider the construction of the distribution $\mathcal{D}$ and partition $\mathcal{P}$ as in \cref{def:tilde_variables_general}. For every $P_i \in \mathcal{P}$, define $$\alpha_0(P_i) = \frac{\mathbb{P}[X_0 \in P_i]}{\mathbb{P}[X_0 \in P_i] + \mathbb{P}[X_1 \in P_i]} ~~ \text{ and } ~~ \alpha_1(P_i) = \frac{\mathbb{P}[X_1 \in P_i]}{\mathbb{P}[X_0 \in P_i] + \mathbb{P}[X_1 \in P_i]}.$$
Then $\mathcal{D}|_{P_i} = \alpha_0(P_i) X_{0}|_{P_i} + \alpha_1(P_i) X_{1}|_{P_i}$.
\end{lemma}
\begin{proof}
    For $x \in P_i$, $$\mathbb{P}_{\mathcal{D}|_{P_i}}[x] = \frac{\mathbb{P}_{\mathcal{D}}[x]}{\mathbb{P}_{\mathcal{D}}[P_i]} = \frac{\frac{1}{2} \mathbb{P}[X_0 = x] + \frac{1}{2} \mathbb{P}[X_1 = x]}{\frac{1}{2} \mathbb{P}[X_0 \in P_i] + \frac{1}{2} \mathbb{P}[X_1 \in P_i]} = \frac{\mathbb{P}[X_0 = x] + \mathbb{P}[X_1 = x]}{\mathbb{P}[X_0 \in P_i] + \mathbb{P}[X_1 \in P_i]}.$$
Relating this to the conditional distributions $X_{0}|_{P_i}$ and $X_{1}|_{P_i}$ of $X_0$ and $X_1$, this expression equals:
$$= \frac{1}{\mathbb{P}[X_0 \in P_i] + \mathbb{P}[X_1 \in P_i]} \cdot \left( \mathbb{P}[X_{0}|_{P_i} = x] \cdot \mathbb{P}[X_0 \in P_i] + \mathbb{P}[X_{1}|_{P_i} = x] \cdot \mathbb{P}[X_1 \in P_i] \right).$$
Given the definition of $\alpha_0(P_i)$ and $\alpha_1(P_i)$, we see that 
\begin{equation}\label{eq:relate-D-to-X0X1}
\mathbb{P}_{\mathcal{D}|_{P_i}}[x] = \alpha_0(P_i) \cdot \mathbb{P}[X_{0}|_{P_i} = x] + \alpha_1(P_i) \cdot \mathbb{P}[X_{1}|_{P_i} = x].
\end{equation}
Stated more concisely, $\mathcal{D}|_{P_i} = \alpha_0(P_i) X_{0}|_{P_i} + \alpha_1(P_i) X_{1}|_{P_i}$. Note that $\alpha_0(P_i) \geq 0$, $\alpha_1(P_i) \geq 0$, and $\alpha_0(P_i) + \alpha_1(P_i) = 1$, so this is indeed a convex combination.
\end{proof}

Next we prove that $X_0$ is computationally indistinguishable from $\widetilde{X}_0$ and $X_1$ is computationally indistinguishable from $\widetilde{X}_1$ on every part of the partition $\mathcal{P}$ guaranteed by applying the Multicalibration Theorem to $g$. We begin by noting that $X_{1}|_{P_i}$ is $(\mathcal{F}, \frac{\eps}{\alpha_1(P_i) (1-\alpha_1(P_i))})$-indistinguishable from $X_{0}|_{P_i}$, which is a simple consequence of the following observation.
\\\\
\noindent \textbf{Observation.} Note that, for all $P_i \in \mathcal{P}$, $v_{P_i} = \alpha_1(P_i)$, which can be seen as follows:
$$v_{P_i} = \mathbb{E}_{x \sim \mathcal{D}|_{P_i}}[g(x)] = \sum_{x \in P_i} \mathbb{P}_{x \sim \mathcal{D}|_{P_i}}[x] \cdot \frac{\mathbb{P}[X_1 = x]}{\mathbb{P}[X_0 = x] + \mathbb{P}[X_1 = x]} $$
$$= \sum_{x \in P_i} \frac{\mathbb{P}_{x \sim \mathcal{D}}[x]}{\mathbb{P}_{\mathcal{D}}[P_i]} \cdot \frac{\mathbb{P}[X_1 = x]}{\mathbb{P}[X_0 = x] + \mathbb{P}[X_1 = x]}$$
$$= \frac{1}{2} \frac{1}{\mathbb{P}_{\mathcal{D}}[P_i]} \sum_{x \in P_i} \mathbb{P}[X_1 = x] = \frac{1}{2} \frac{1}{\frac{1}{2}\mathbb{P}[X_0 \in P_i] + \frac{1}{2}\mathbb{P}[X_1 \in P_i]} \mathbb{P}[X_1 \in P_i]$$
$$= \frac{\mathbb{P}[X_1 \in P_i]}{\mathbb{P}[X_0 \in P_i] + \mathbb{P}[X_1 \in P_i]}.$$

\begin{lemma}\label{lem:X_0-and-tilde-on-partition}
    For random variables $X_0, X_1$ and family of functions $\mathcal{F}$, consider the construction of the family of functions $\mathcal{F}'$, the distribution $\mathcal{D}$, function $g$, and partition $\mathcal{P}$ as in \cref{def:tilde_variables_general}. For $P_i \in \mathcal{P}$, consider the definitions of $\alpha_0(P_i)$ and $\alpha_1(P_i)$ as in \cref{lem:D-and-X0X1}.
    
    For all $P_i \in \mathcal{P}$ such that $\mathbb{P}_{x \sim \mathcal{D}}[x \in P_i] \geq \gamma$, $X_{0}|_{P_i}$ is $(\mathcal{F}, \frac{\eps}{1-\alpha_1(P_i)})$-indistinguishable from $\widetilde{X}_{0}|_{P_i}$ and $X_{1}|_{P_i}$ is $(\mathcal{F}, \frac{\eps}{\alpha_1(P_i)})$-indistinguishable from $\widetilde{X}_{1}|_{P_i}$.
\end{lemma}

The idea behind the proof of this lemma is as follows. First, since for all $P_i \in \mathcal{P}$, $\widetilde{X}_{0}|_{P_i}$ is equivalent to $\mathcal{D}|_{P_i}$, we want to argue that $\mathcal{D}|_{P_i}$ must also be indistinguishable from $X_{0}|_{P_i}$ on all $P_i \in \mathcal{P}$ such that $\mathbb{P}_{x \sim \mathcal{D}}[x \in P_i] \geq \gamma$. This is implied by the facts that $\mathcal{D}|_{P_i}$ is a convex combination of $X_{0}|_{P_i}$ and $X_{1}|_{P_i}$ and these two random variables are indistinguishable for these $P_i \in \mathcal{P}$.

\begin{proof}
    By definition of $\widetilde{X}_0$, for $x \in P_i$:
    $$\mathbb{P}[\widetilde{X}_0 = x] = \mathbb{P}[X_0 \in P_i] \cdot \mathbb{P}_{\mathcal{D}|_{P_i}}[x].$$
    From Lemma \ref{lem:D-and-X0X1}, this equals:
    $$= \mathbb{P}[X_0 \in P_i] \cdot \left(\alpha_0(P_i) \cdot \mathbb{P}[X_{0}|_{P_i} = x] + \alpha_1(P_i) \cdot \mathbb{P}[X_{1}|_{P_i} = x] \right).$$
    Define $\widetilde{X}_{0}|_{P_i}$ to be the random variable such that for $x \in P_i$, $\mathbb{P}[\widetilde{X}_{0}|_{P_i} = x] = \mathbb{P}[\widetilde{X}_{0} = x | x \in P_i]$. Note that this equals $\mathbb{P}_{\mathcal{D}|_{P_i}}[x]$. Recall that, from Lemma \ref{lem:X0Pi_vs_X1Pi}, for all  $P_i \in \mathcal{P}$ such that $\mathbb{P}_{x \sim \mathcal{D}}[x \in P_i] \geq \gamma$ $X_{1}|_{P_i}$ is $(\mathcal{F}, \frac{\eps}{\alpha_1(P_i) (1-\alpha_1(P_i))})$-indistinguishable from $X_{0}|_{P_i}$.

   To show that $X_{0}|_{P_i}$ is $(\mathcal{F}, \frac{\eps}{1-\alpha_1(P_i)})$-indistinguishable from $\widetilde{X}_{0}|_{P_i}$ we need to bound, for every $f \in \mathcal{F}'$:
   $\left| \sum_{x \in P_i} f(x) \cdot \left( \mathbb{P}_{\mathcal{D}|_{P_i}}[x] - \mathbb{P}_{X_{0}|_{P_i}}[x]\right)\right|.$

   We see that 
   $$\left| \sum_{x \in P_i} f(x) \cdot \left( \mathbb{P}_{\mathcal{D}|_{P_i}}[x] - \mathbb{P}_{X_{0}|_{P_i}}[x]\right)\right| $$$$
   = \left| \sum_{x \in P_i} f(x) \cdot \left( (\alpha_0(P_i) - 1) \mathbb{P}_{X_{0}|_{P_i}}[x] + \alpha_1(P_i) \mathbb{P}_{X_{1}|_{P_i}}[x]\right)\right|$$
   $$= \alpha_1(P_i) \left| \sum_{x \in P_i} f(x) \cdot  \left( \mathbb{P}_{X_{0}|_{P_i}}[x] - \mathbb{P}_{X_{1}|_{P_i}}[x]\right)\right| \leq \frac{\eps}{1 - \alpha_1(P_i)}.$$
   Similarly, we can also show that $X_{1}|_{P_i}$ is $(\mathcal{F}, \frac{\eps}{\alpha_1(P_i)})$-indistinguishable from $\widetilde{X}_{1}|_{P_i}$.
   \end{proof}

We are finally ready to prove \Cref{thm:MCproperties_restatement} part (a).

\begin{proof}

For random variables $X_0, X_1$ and family of functions $\mathcal{F}$, consider the construction of the family of functions $\mathcal{F}'$, the distribution $\mathcal{D}$, function $g$, and partition $\mathcal{P}$ as in \cref{def:tilde_variables_general}.

We use the analysis of indistinguishability of random variables over all $P_i \in \mathcal{P}$ such that $\mathbb{P}_{x \sim \mathcal{D}}[x \in P_i] \geq \gamma$ to prove the indistinguishability of $X_0$ and $\widetilde{X}_0$ globally over the domain. Because we have results about indistinguishability over parts of the partition that have enough weight with respect to $\mathcal{D}$, but not those whose weight is too small, we need to break up the analysis to handle both types of $P_i \in \mathcal{P}$. Let $\mathcal{P}(\gamma)$ be the set of $P_i \in \mathcal{P}$ such that $\mathbb{P}_{x \sim \mathcal{D}}[x \in P_i] \geq \gamma$.

We focus on the indistinguishability of $X_0$ from $\widetilde{X}_0$ in the proof. The proof of indistinguishability of $X_1$ from $\widetilde{X}_1$ follows by similar arguments.

$$\left|\sum_{P_i \in \mathcal{P}} \sum_{x \in P_i} f(x) \cdot \left( \mathbb{P}[X_0 = x] -  \mathbb{P}[\widetilde{X}_0 = x]\right) \right| $$
$$\leq \sum_{P_i \in \mathcal{P}(\gamma)} \left| \sum_{x \in P_i} f(x) \cdot \left( \mathbb{P}[X_0 = x] -  \mathbb{P}[\widetilde{X}_0 = x]\right) \right| $$
\begin{equation}\label{eq:lemma_X0_and_tilde_equation}
+ \sum_{P_i \in \mathcal{P} \setminus \mathcal{P}(\gamma)} \left|\sum_{x \in P_i} f(x) \cdot \left( \mathbb{P}[X_0 = x] -  \mathbb{P}[\widetilde{X}_0 = x]\right) \right|.
\end{equation}
Let us focus on the first of the two summations in Equation (\ref{eq:lemma_X0_and_tilde_equation}). We see 
$$\sum_{P_i \in \mathcal{P}(\gamma)} \left| \sum_{x \in P_i} f(x) \cdot \left( \mathbb{P}[X_0 = x] -  \mathbb{P}[\widetilde{X}_0 = x]\right) \right|$$
$$= \sum_{P_i \in \mathcal{P}(\gamma)} \mathbb{P}[X_0 \in P_i]  \left| \sum_{x \in P_i} f(x) \cdot \left( \mathbb{P}[X_{0}|_{P_i} = x] -  \mathbb{P}[\widetilde{X}_{0}|_{P_i} = x]\right) \right|.$$
Applying Lemma \ref{lem:X_0-and-tilde-on-partition}, this is:
$$\leq \sum_{P_i \in \mathcal{P}(\gamma)} \mathbb{P}[X_0 \in P_i] \cdot \frac{\eps}{\alpha_0 (P_i)} = \sum_{P_i \in \mathcal{P}(\gamma)} \alpha_0(P_i) \cdot 2 \mathbb{P}[\mathcal{D} \in P_i] \frac{\eps}{\alpha_0 (P_i)} = 2 \eps.$$

Let us now focus on the second of the two summations in Equation (\ref{eq:lemma_X0_and_tilde_equation}). We see 
$$\sum_{P_i \in \mathcal{P} \setminus \mathcal{P}(\gamma)} \left|\sum_{x \in P_i} f(x) \cdot \left( \mathbb{P}[X_0 = x] -  \mathbb{P}[\widetilde{X}_0 = x]\right) \right| \leq \sum_{P_i \in \mathcal{P} \setminus \mathcal{P}(\gamma)} \left(\mathbb{P}[X_0 \in P_i] + \mathbb{P}[\widetilde{X}_0 \in P_i] \right)$$ $$= \sum_{P_i \in \mathcal{P} \setminus \mathcal{P}(\gamma)} 2 \mathbb{P}[X_0 \in P_i] < 4 \gamma |\mathcal{P}|,$$
where the last inequality follows from observing that $\frac{1}{2} \mathbb{P}[X_0 \in P_i] \leq \mathbb{P}[\mathcal{D} \in P_i] < \gamma$.  Note that, by definition of $\gamma$ and by the bounds on $|\mathcal{P}|$ from the Multicalibration Theorem, $4 \gamma |\mathcal{P}|$ is $O(\eps)$.

    Combining the different components of the analysis above, we find that:
    $$\left|\sum_{P_i \in \mathcal{P}} \sum_{x \in P_i} f(x) \cdot \left( \mathbb{P}[X_0 = x] -  \mathbb{P}[\widetilde{X}_0 = x]\right) \right| \leq O(\eps).$$ Therefore, $X_0$ is $(\mathcal{F}, O(\eps))$-indistinguishable from $\widetilde{X}_0$. Similarly, $X_1$ is $(\mathcal{F}, O(\eps))$-indistinguishable from $\widetilde{X}_1$.

    Additionally, to conclude the theorem, note that by definition of $\widetilde{X}_0, \widetilde{X}_1$, for the partition $\mathcal{P}$ given by the Multicalibration Theorem that $\widetilde{X}_0, \widetilde{X}_1$ are defined according to, for all $P_i \in \mathcal{P}$, $\widetilde{X}_{0}|_{P_i} = \widetilde{X}_{1}|_{P_i}$.
\end{proof}

\subsection{Proof of Parts (b, c, d)}

Here, we prove the remaining properties of \cref{thm:MCproperties_restatement}: 

\begin{proof}
    For random variables $X_0, X_1$ and family of functions $\mathcal{F}$, consider the construction of the family of functions $\mathcal{F}'$, the distribution $\mathcal{D}$, function $g$, partition $\mathcal{P}$, and function $p: \mathcal{X} \to [m]$ as in \cref{def:tilde_variables_general}. We show the following:
\begin{claim}[Part (b) of \cref{thm:MCproperties_restatement}]
            $p(X_0)$ is identically distributed to $p(\widetilde{X}_0)$ and $p(X_1)$ is identically distributed to $p(\widetilde{X}_1)$.
            \end{claim}
            \begin{proof}
                We show this WLOG for $X_0, \widetilde{X}_0$. This follows by definition. Recall that to construct $\widetilde{X}_0$, we sample part $P_i$ with probability $\Pr[X_0 \in P_i]$, and then replace the conditional distribution over $P_i$ with $\mathcal{D}$ (as defined in \cref{def:tilde_variables_general}). Thus, for any $P_i \in \mathcal{P}$, we have $\Pr[\widetilde{X}_0 \in P_i] = \Pr[X_0 \in P_i]$.
            \end{proof}
\begin{claim}[Part (c) of \cref{thm:MCproperties_restatement}]
    For every in $y \in [m]$, $\widetilde{X}_{0|p(\widetilde{X}_0)=y}$ is identically distributed to $\widetilde{X}_{1|p(\widetilde{X}_1)=y}$.
\end{claim}
\begin{proof}
    Again, this is merely definitional. As in \cref{def:tilde_variables_general}, for any part $P_i \in \mathcal{P}$, we define 
    \[
    \widetilde{X}_{1|P_i} = \mathcal{D}_{P_i},
    \]
    and likewise 
    \[
    \widetilde{X}_{0|P_i} = \mathcal{D}_{P_i}.
    \]
    Thus, the two distributions have the same marginal when conditioned on being in any part $P_i$. We conclude by recalling that the parts $P_i$ are exactly the pieces $P^{-1}(y)$, for $y \in [m]$, hence yielding the statement. 
\end{proof}
\begin{claim}[Part (d) of \cref{thm:MCproperties_restatement}]
    $p$ is computable by circuits in $\mathcal{F}_{O(1 / \eps^6), O(1 / (\eps^{12}) \cdot \log(|\mathcal{X}|)\log(|\mathcal{X}| / \eps) )}$.
\end{claim}

\begin{proof}
    Recall that in \cref{def:tilde_variables_general}, $p$ is exactly the partition function that results from constructing a multicalibrated partition with parameters $\eps, \gamma = \eps^2$ on the family $\mathcal{F}'$ (where $\mathcal{F}' = \mathcal{F}_{c, c \log(|\mathcal{X}|)}$). As in \cref{thm:multicalibrationGeneralFunctions}, such a partition function can be computed by a function in 
    \[
    \mathcal{F}'_{O(1 / \eps^6), O(1 / (\eps^{12}) \cdot \log(|\mathcal{X}| / \eps) )}.
    \]
    This yields the claim. 
\end{proof}

The proof of \cref{thm:MCproperties_restatement} then follows from the union of each of the individual claims.
\end{proof}

\section{Proof of \cref{thm:main}}\label{sec:mainLB}

We now prove that the computational indistinguishability of $X_0^{\otimes k}$ and $X_1^{\otimes k}$ is captured by the total variation distance between $\widetilde{X}_0^{\otimes k}$ and $\widetilde{X}_1^{\otimes k}$, which is \Cref{thm:main} (2). Before stating the general version of the theorem that we prove, we define the types of families of functions that our theorem applies to, which are \textit{closed under marginals}. 

\begin{definition}\label{def:marginals}
    Consider a family of functions $\mathcal{F}$ over $\mathcal{X}$. We define a new family of functions $\mathcal{F}^{\otimes k}$ over $\mathcal{X}^{k}$ to be the set of all functions $f:\mathcal{F}^{\otimes k} \rightarrow [0,1]$ such that 
    for every $f \in \mathcal{F}^{\otimes k}$, 
    for every $j \in [k]$, and every fixing of the variables $(x_i)_{i \in [k]\setminus j}$, it is the case that 
    \[
    f(x_1, x_2, \dots, x_{j-1}, \cdot , x_{j+1}, \dots x_k) = f'(\cdot),
    \]
    for some function $f'\in \mathcal{F}$.
\end{definition}

\textbf{Observation.} Notice that when $f: \mathcal{X}^k \rightarrow [0,1]$ is a function computable by a size $s$ circuit, then every marginal is also computable by a size $s$ circuit.

Given this definition, we are now ready to state the general versions of the theorems we prove.

\begin{theorem}\label{thm:mainRestated}
    For every pair of random variables $X_0, X_1$ over $\mathcal{X}$, every family of functions $\mathcal{F}$, and every $\eps > 0$, there exist random variables $\widetilde{X}_0, \widetilde{X}_1$ such that for every $k>0$,
    \begin{enumerate}
        \item $X_b$ is $\eps$-indistinguishable from $\widetilde{X}_b$ by functions in $\mathcal{F}^{\otimes k}$, for each $b\in \zo$.
        \item $X_0^{\otimes k}$ and $X_1^{\otimes k}$ are $(d_{TV}(\widetilde{X}_0^{\otimes k}, \widetilde{X}_1^{\otimes k})+2k \cdot \eps)$-indistinguishable by functions in $\mathcal{F}^{\otimes k}$.
        \item $X_0^{\otimes k}$ and $X_1^{\otimes k}$ are $(d_{TV}( \widetilde{X}_0^{\otimes k}, \widetilde{X}_1^{\otimes k})- 2k \cdot \eps)$-distinguishable by functions in $\mathcal{F}_{K_2,K_1}$, where $K_1 = O(\frac{k}{\eps^{12}} \cdot \log(|\mathcal{X}|) \cdot \log(|\mathcal{X}|/\eps)) + O(k \cdot \log(|\mathcal{X}|) \cdot \polylog(k / \eps))$ and $K_2 = O(k/\eps^6)$.
    \end{enumerate}
\end{theorem}

\begin{proof}
    Apply \cref{thm:MCproperties_restatement} to random variables $X_0, X_1$ and the family $\mathcal{F}$ of functions to obtain random variables $\widetilde{X}_0, \widetilde{X}_1$ and function $p: \mathcal{X} \to [m]$. Then, Item (1) follows immediately from \cref{thm:MCproperties_restatement}. Item (2) follows from \cref{lemma:computational-indistinguishability-bound}, proven below. Item (3) follows from \cref{thm:distinguishingCircuit}, proven below, when setting $\delta = \eps / 100k$.
\end{proof}

In the remaining sections, we provide the proofs and preliminaries for \cref{lemma:computational-indistinguishability-bound} and \cref{thm:distinguishingCircuit}.

\subsection{Indistinguishability of $X_b^{\otimes k}$ and $\widetilde{X}_b^{\otimes k}$}

We want to bound the computational indistinguishability of $X_0^{\otimes k}$ and $X_1^{\otimes k}$. First, let's use the arguments from the previous subsection to prove the following.

\begin{lemma}\label{lem:moving-to-product-distributions}
    Let $\mathcal{F}$ be a family of functions mapping from $\mathcal{X} \rightarrow [0,1]$. 
    For random variables $X_0, X_1$ and family $\mathcal{F}$, consider the construction of random variables $\widetilde{X}_0, \widetilde{X}_1$ as promised by \cref{thm:MCproperties_restatement}. Then, for $f \in \mathcal{F}^{\otimes k}$ and $i \in \{0, 1\}$, 
    $$\left| \sum_{\Bar{x} \in \mathcal{X}^{k}} f(\Bar{x}) \cdot \left(\mathbb{P}[X_{i}^{\otimes k} = \Bar{x}] - \mathbb{P}[\widetilde{X}_{i}^{\otimes k} = \Bar{x}] \right) \right| \leq O(k \eps).$$
\end{lemma}

\begin{proof}[Proof of Lemma \ref{lem:moving-to-product-distributions}]
We proceed by a standard hybrid argument.

    Consider $i = 0$; the case of $i = 1$ follows similarly. Let $X_0^{\otimes (k - j)} \widetilde{X}_0^{\otimes j}$ denote a product distribution such that the first $k - j$ coordinates are distributed according to random variable $X_0$ and the remaining $j$ coordinates are distributed according to $\widetilde{X}_0$. Note that:
    $$\mathbb{P}[X_{0}^{\otimes k} = \Bar{x}] - \mathbb{P}[\widetilde{X}_{0}^{\otimes k} = \Bar{x}] = \sum_{j = 0}^{k-1} \left(\mathbb{P}[X_{0}^{\otimes (k-j)}\widetilde{X}_0^{\otimes j} = \Bar{x}] - \mathbb{P}[X_{0}^{\otimes (k-j-1)}\widetilde{X}_0^{\otimes (j+1)} = \Bar{x}] \right).$$
    Let $\Bar{x}_i$ be the value that $\Bar{x}$ takes in the $i$th coordinate. Splitting up the probability in the previous expression into various terms, the expression equals:
    $$= \sum_{j = 0}^{k-1} \mathbb{P}[X_0^{\otimes (k-j-1)} = \bar{x}_{1} \bar{x}_2 \dots \bar{x}_{k - j - 1}] \cdot \mathbb{P}[\widetilde{X}_0^{\otimes j} = \bar{x}_{k - j + 1} \bar{x}_{k - j + 2} \dots \bar{x}_{k - j - 1}] \cdot \left( \mathbb{P}[X_0 = \Bar{x}_{k - j}] - \mathbb{P}[\widetilde{X}_0 = \Bar{x}_{k - j}]\right).$$

    Applying this and the triangle inequality, we find:
    $$\left| \sum_{\Bar{x} \in \mathcal{X}^{k}} f(\Bar{x}) \cdot \left(\mathbb{P}[X_{i}^{\otimes k} = \Bar{x}] - \mathbb{P}[\widetilde{X}_{i}^{\otimes k} = \Bar{x}] \right) \right| $$ $$\leq \sum_{j = 0}^{k-1} \bigg| \sum_{\Bar{x} \in \mathcal{X}^{k}} f(\Bar{x}) \cdot \mathbb{P}[X_0^{\otimes (k-j-1)} = \bar{x}_{1} \bar{x}_2 \dots \bar{x}_{k - j - 1}] \cdot \mathbb{P}[\widetilde{X}_0^{\otimes j} = \bar{x}_{k - j + 1} \bar{x}_{k - j + 2} \dots \bar{x}_{k}] $$ $$\cdot \left( \mathbb{P}[X_0 = \Bar{x}_{k - j}] - \mathbb{P}[\widetilde{X}_0 = \Bar{x}_{k - j}]\right)  \bigg|$$
    $$\leq \sum_{j = 0}^{k - 1}  \mathbb{E}_{\bar{x}_{1} \bar{x}_2 \dots \bar{x}_{k - j - 1}\bar{x}_{k - j + 1} \bar{x}_{k - j + 2} \dots \bar{x}_{k} \sim X_0^{\otimes (k-j-1)} \widetilde{X}_0^{\otimes j}} \bigg[ \bigg|\mathbb{E}_{a \sim X_0}\big[f((\bar{x}_1, \bar{x}_2, \dots, \bar{x}_{k - j - 1}, a, \bar{x}_{k - j + 1}, \bar{x}_{k - j + 2}, \dots, \bar{x}_k))\big] $$ $$- \mathbb{E}_{a \sim \widetilde{X}_0}\big[f((\bar{x}_1, \bar{x}_2, \dots, \bar{x}_{k - j - 1}, a, \bar{x}_{k - j + 1}, \bar{x}_{k - j + 2}, \dots, \bar{x}_k))\big] \bigg|\bigg]$$

    Since we assume that the marginal of $\mathcal{F}$ is also in $\mathcal{F}$ there exists some $h_{\bar{x}_{1} \bar{x}_2 \dots \bar{x}_{k - j - 1}\bar{x}_{k - j + 1} \bar{x}_{k - j + 2} \dots \bar{x}_{k}} \in \mathcal{F}$ such that $f((\bar{x}_1, \bar{x}_2, \dots, \bar{x}_{k - j - 1}, \cdot, \bar{x}_{k - j + 1}, \bar{x}_{k - j + 2}, \dots, \bar{x}_k)) = h_{\bar{x}_{1} \bar{x}_2 \dots \bar{x}_{k - j - 1}\bar{x}_{k - j + 1} \bar{x}_{k - j + 2} \dots \bar{x}_{k}}(\cdot)$.
    Given this requirement, the inequality above equals:
    $$= \sum_{j = 0}^{k - 1}  \mathbb{E}_{\bar{x}_{1} \bar{x}_2 \dots \bar{x}_{k - j - 1}\bar{x}_{k - j + 1} \bar{x}_{k - j + 2} \dots \bar{x}_{k} \sim X_0^{\otimes (k-j-1)} \widetilde{X}_0^{\otimes j}} \bigg[ \bigg|\mathbb{E}_{a \sim X_0}\big[h_{\bar{x}_{1} \bar{x}_2 \dots \bar{x}_{k - j - 1}\bar{x}_{k - j + 1} \bar{x}_{k - j + 2} \dots \bar{x}_{k}}(a)\big] $$ $$- \mathbb{E}_{a \sim \widetilde{X}_0}\big[h_{\bar{x}_{1} \bar{x}_2 \dots \bar{x}_{k - j - 1}\bar{x}_{k - j + 1} \bar{x}_{k - j + 2} \dots \bar{x}_{k}}(a)\big] \bigg|\bigg].$$
    Using \cref{thm:MCproperties_restatement}, for some constant $C$, this is:
    $$\leq \sum_{j = 0}^{k - 1}  \mathbb{E}_{\bar{x}_{1} \bar{x}_2 \dots \bar{x}_{k - j - 1}\bar{x}_{k - j + 1} \bar{x}_{k - j + 2} \dots \bar{x}_{k} \sim X_0^{\otimes (k-j-1)} \widetilde{X}_0^{\otimes j}} \bigg[ C \eps\bigg] = O(k \eps). \eqno \qedhere$$
\end{proof}

\subsection{Proving Item (2) of \cref{thm:mainRestated}}\label{subsec:hellinger}

Recall that the computational indistinguishability of  $\widetilde{X}_0^{\otimes k}$ and $\widetilde{X}_1^{\otimes k}$ is implied by the statistical indistinguishability of the two random variables, given by the following remark.

\begin{remark}\label{remark:computational_statistical}
    Let $\mathcal{F}$ be a family of functions mapping from $\mathcal{X} \rightarrow [0,1]$. 
    For random variables $X_0, X_1$ and family $\mathcal{F}$, consider the construction of random variables $\widetilde{X}_0, \widetilde{X}_1$ as promised by \cref{thm:MCproperties_restatement}. Consider $f \in \mathcal{F}^{\otimes k}$. Then $$\left|\sum_{\Bar{P} \in \mathcal{P}^{\otimes k}} \sum_{\Bar{x} \in \Bar{P}} f(x) \cdot \left( \mathbb{P}[\widetilde{X}_0^{\otimes k} = \Bar{x}] - \mathbb{P}[\widetilde{X}_1^{\otimes k} = \Bar{x}]\right) \right| \leq d_{\mathrm{TV}}(\widetilde{X}_0^{\otimes k}, \widetilde{X}_1^{\otimes k}).$$
\end{remark}

Combining this remark with the results of the previous subsections, we obtain the following key lemma:

\begin{lemma}\label{lemma:computational-indistinguishability-bound}
    For random variables $X_0, X_1$ and family $\mathcal{F}$ of functions, consider the construction of random variables $\widetilde{X}_0, \widetilde{X}_1$ as promised by \cref{thm:MCproperties_restatement}. For all $f \in \mathcal{F}^{\otimes k}$, the following three inequalities hold:
$$\left| \sum_{\Bar{x} \in \mathcal{X}^{k}} f(\Bar{x}) \cdot \left(\mathbb{P}[X_{0}^{\otimes k} = \Bar{x}] - \mathbb{P}[X_{1}^{\otimes k} = \Bar{x}] \right) \right| \leq O(k \eps) + d_{\mathrm{TV}}(\widetilde{X}_0^{\otimes k}, \widetilde{X}_1^{\otimes k})$$ $$\leq O(k \eps) + \sqrt{2} \sqrt{1 - (1 - d_H^2(\widetilde{X}_0, \widetilde{X}_1))^k}$$ $$\leq  O(k \eps)  + \sqrt{2k} \cdot d_H(\widetilde{X}_0, \widetilde{X}_1).$$
\end{lemma}

\begin{proof}
    For $f \in \mathcal{F}^{\otimes k}$,
    \begin{eqnarray}
\nonumber\lefteqn{\left| \sum_{\Bar{x} \in \mathcal{X}^{k}} f(\Bar{x}) \cdot \left(\mathbb{P}[X_{0}^{\otimes k} = x] - \mathbb{P}[X_{1}^{\otimes k} = x] \right) \right| } \\
    \nonumber & & \leq \left| \sum_{\Bar{x} \in \mathcal{X}^{k}} f(\Bar{x}) \cdot \left(\mathbb{P}[X_{0}^{\otimes k} = \Bar{x}] - \mathbb{P}[\widetilde{X}_0^{\otimes k} = \Bar{x}] \right) \right|\\
    \nonumber & & + \left| \sum_{\Bar{x} \in \mathcal{X}^{k}} f(\Bar{x}) \cdot \left(\mathbb{P}[\widetilde{X}_0^{\otimes k} = \Bar{x}] - \mathbb{P}[\widetilde{X}_1^{\otimes k} = \Bar{x}] \right) \right| \\
    & & + \left| \sum_{\Bar{x} \in \mathcal{X}^{k}} f(\Bar{x}) \cdot \left(\mathbb{P}[\widetilde{X}_1^{\otimes k} = \Bar{x}] - \mathbb{P}[X_{1}^{\otimes k} = \Bar{x}] \right) \right|.
    \end{eqnarray}
From the previous sections and the remark above, this is:
$$\leq O(k \eps) + d_{\mathrm{TV}}(\widetilde{X}_0^{\otimes k}, \widetilde{X}_1^{\otimes k}).$$
We relate this to the Hellinger distance $d_H$ between $\widetilde{X}_0^{\otimes k}$ and $\widetilde{X}_1^{\otimes k}$:
$$\leq O(k \eps)  + \sqrt{2} d_{H}(\widetilde{X}_0^{\otimes k}, \widetilde{X}_1^{\otimes k}).$$
Using the fact that $1 -d_H^2(\widetilde{X}_0^{\otimes k}, \widetilde{X}_1^{\otimes k}) = (1 - d_H^2(\widetilde{X}_0, \widetilde{X}_1))^k$, the expression above is:
$$ = O(k \eps) + \sqrt{2} \sqrt{1 - ( 1 - d_H^2(\widetilde{X}_0, \widetilde{X}_1))^k}.$$
Note that by Claim \ref{claim:hellinger_facts} this is:
$$\leq  O(k \eps)  + \sqrt{2k} \cdot d_H(\widetilde{X}_0, \widetilde{X}_1). \eqno \qedhere$$
\end{proof}

\subsection{Small Circuits that Distinguish Between $X_0^{\otimes k}, X_1^{\otimes k}$: Proof of Item (3)}\label{sec:efficient}

In this section, we prove item (3) of \cref{thm:mainRestated} (stated as \cref{thm:distinguishingCircuit}). We state this below:

\begin{lemma}\label{thm:distinguishingCircuit}
    Let $X_0, X_1$ be random variables, and let $k$ be an integer. For random variables $X_0, X_1$, consider the construction of random variables $\widetilde{X}_0, \widetilde{X}_1$ as promised in \cref{thm:MCproperties_restatement} with respect to a class of functions $\mathcal{F}$. Then, there exists a circuit in 
    \[
     \mathcal{F}_{K_2, K_1}
    \]
    with $K_1 = k/\eps^{12} \cdot \log(|\mathcal{X}|) \cdot\log(|\mathcal{X}|/\eps ) + O(k \cdot \polylog(1 /(\eps \delta)))$, and $K_2 = O(k/\eps^6)$
    which distinguishes $X^{\otimes k}_0, X^{\otimes k}_1$ with advantage $d_{\mathrm{TV}}(\widetilde{X}_{0}^{\otimes k}, \widetilde{X}_{1}^{\otimes k}) - 4\delta k$.
\end{lemma}

\begin{proof}[Proof of \cref{thm:distinguishingCircuit}]
For random variables $X_0, X_1$ and family of functions $\mathcal{F}$, consider the random variables $\widetilde{X}_0, \widetilde{X}_1$, and partition function $p$ as defined in \cref{thm:MCproperties_restatement}. Recall that by this theorem, the partitions $\mathcal{P} \in \mathcal{F}_{q,t}$, with $t = O(1 /(\eps^6) \cdot \log(|\mathcal{X}|)/\eps ), q = O(1 / \eps^6)$, in the sense that there is a circuit of the above stated complexity which can compute the index of the partition that each element is in. 

Now, for $P_i \in \mathcal{P}$ let 
\[
\alpha_0(P_i) = \frac{\Pr[X_0 \in P_i]}{\Pr[X_0 \in P_i] + \Pr[X_1 \in P_i]},
\]
and define $\alpha_1(P_i)$ similarly. 

Next, let us define new distributions $\widetilde{X}_{0, \delta}, \widetilde{X}_{1, \delta}$, where all of the $\alpha(P_i)$ values are arbitrarily shifted to a neighboring multiple of $\delta/r$. That is to say if $\alpha_0(P_i) \in [\ell \delta / r, (\ell+1)\delta / r]$ for an integer $\ell$, then we arbitrarily set $\alpha_{0, \delta}(P_i)$ to be one of $\ell \delta / r, (\ell+1)\delta / r$.

Observe that 
\[
d_{\mathrm{TV}}(\widetilde{X}_{0,\delta}, \widetilde{X}_0) \leq \delta,
\]
as there are $r$ parts in $\mathcal{P}$, and each has its probability mass shifted by at most $\delta / r$. Note that this also means that 
\[
d_{\mathrm{TV}}(\widetilde{X}_{0, \delta}^{\otimes k}, \widetilde{X}_0^{\otimes k}) \leq \delta k.
\]

Now, our goal will be to construct an optimal distinguisher for $\widetilde{X}_{0, \delta}^{\otimes k}$ and $\widetilde{X}_{1, \delta}^{\otimes k}$. To do this, consider a sequence of samples $x_1, \dots x_k$, where $x_1 \in P_{i_1}, \dots , x_k \in P_{i_k}$. Observe that we can exactly compute the relative probability that $x_1, \dots x_k$ comes from $\widetilde{X}_{0, \delta}^{\otimes k}$ vs. $\widetilde{X}_{1, \delta}^{\otimes k}$. 

Indeed, let 
\[
\kappa((x_1, \dots x_k)) = \frac{\Pr[\widetilde{X}_{1, \delta}^{\otimes k} = (x_1, \dots x_k)]}{\Pr[\widetilde{X}_{0, \delta}^{\otimes k} = (x_1, \dots x_k)]} = \prod_{j = 1}^k \frac{\alpha_{1, \delta}(P_{i_j})}{\alpha_{0, \delta}(P_{i_j})}.
\]

For a sequence of $k$ samples $x_1, \dots x_k$, the optimal distinguisher is simply the distinguisher which returns $1$ (corresponding to deciding that the samples are from $\widetilde{X}_{1, \delta}^{\otimes k}$) if the above expression is $\geq 1$, and $0$ otherwise (corresponding to deciding that the samples are from $\widetilde{X}_{0, \delta}^{\otimes k}$). This is because in general for two distributions $p, q$, we have that 
\[
d_{\mathrm{TV}}(p, q) = \sum_{x \in \mathcal{X}} (\max(p(x), q(x)) - \min(p(x), q(x))).
\]

That is to say, if we let $C((x_1, \dots x_k))$ be the circuit which computes $\mathbf{1}[\kappa((x_1, \dots x_k)) \geq 1]$, we have
\[
\left | \sum_{(x_1, \dots x_k) \in \mathcal{X}^k} C((x_1, \dots x_k)) \cdot \left [\widetilde{X}_{0, \delta}^{\otimes k}(x_1, \dots x_k) - \widetilde{X}_{1, \delta}^{\otimes k}(x_1, \dots x_k) \right ] \right | = d_{\mathrm{TV}}(\widetilde{X}_{0, \delta}^{\otimes k}, \widetilde{X}_{1, \delta}^{\otimes k}).
\]

Now, we will show that we can actually compute such a circuit $C$ which is not too complex. Indeed, the key observation is that because we ``rounded'' our distributions initially, now when we compute the product $\prod_{j = 1}^k \frac{\alpha_{1, \delta}(P_{i_j})}{\alpha_{0, \delta}(P_{i_j})}$, every pair of $\alpha_{1, \delta}(P_{i_j}), \alpha_{0, \delta}(P_{i_j})$ share a denominator. Thus, computing this product is equivalent to multiplying together $k$ integers in $[r / \delta]$. So, we consider the following circuit: first, $C$ has $k$ copies of the circuit which, upon receiving element $x$, computes the index $i$ for which $x \in P_i$. Recall that this circuit is $\in \mathcal{F}_{t, q}$. Next, once we have computed the index $i$ of the partition that an element $x$ is in, we hard-wire two-lookup tables: one computes the numerator of $\alpha_{0, \delta}(P_i)$, and the other computes the numerator of $\alpha_{1, \delta}(P_i)$. Observe that because there are only $r$ parts in the partition, the complexity of these look-up tables is not too large. By \cite{Lup70}, each look-up table requires only circuits of size $O(r)$. Once we have computed the numerators of these values, all that remains is to multiply together these integers, and return the index $b \in \zo$, corresponding to whichever product is larger. Computing this product can be done with circuits of size $O(k \cdot \polylog(r / \delta))$.

Thus, in total, this circuit $C$ for distinguishing $\widetilde{X}_{0, \delta},\widetilde{X}_{1, \delta} $ can be implemented in \[\mathcal{F}_{qk, tk + O(rk) + O(k \cdot \polylog(r / \delta))},
\]
as we use $k$ copies of a circuit in $\mathcal{F}_{q,t}$, along with $O(k)$ look-up tables, and $O(k \cdot \polylog(r / \delta))$ extra wires to compute the product. By plugging in $q = O(1 / \eps^6), r= O(1/\eps),t = O(1 /(\eps^6 \gamma) \cdot \log(|\mathcal{X}|)\cdot\log(|\mathcal{X}|)/\eps )$, we obtain the stated complexity.

Finally, it remains to show that this circuit $C$ does a good job of distinguishing $X_0, X_1$. To this end, let us first observe that 
\[
d_{\mathrm{TV}}(\widetilde{X}_{0, \delta}^{\otimes k}, \widetilde{X}_{1, \delta}^{\otimes k}) \geq d_{\mathrm{TV}}(\widetilde{X}_{0}^{\otimes k}, \widetilde{X}_{1}^{\otimes k}) - d_{\mathrm{TV}}(\widetilde{X}_{0}^{\otimes k}, \widetilde{X}_{0, \delta}^{\otimes k}) - d_{\mathrm{TV}}(\widetilde{X}_{1}^{\otimes k}, \widetilde{X}_{1, \delta}^{\otimes k}).
\]

Thus, $d_{\mathrm{TV}}(\widetilde{X}_{0, \delta}^{\otimes k}, \widetilde{X}_{1, \delta}^{\otimes k}) \geq d_{\mathrm{TV}}(\widetilde{X}_{0}^{\otimes k}, \widetilde{X}_{1}^{\otimes k}) - 2\delta k$. This means that we have
\[
\left | \sum_{(x_1, \dots x_k) \in \mathcal{X}^k} C((x_1, \dots x_k)) \cdot \left [\widetilde{X}_{0, \delta}^{\otimes k}(x_1, \dots x_k) - \widetilde{X}_{1, \delta}^{\otimes k}(x_1, \dots x_k) \right ] \right | 
\]
\[
= d_{\mathrm{TV}}(\widetilde{X}_{0, \delta}^{\otimes k}, \widetilde{X}_{1, \delta}^{\otimes k}) \geq d_{\mathrm{TV}}(\widetilde{X}_{0}^{\otimes k}, \widetilde{X}_{1}^{\otimes k}) - 2\delta k.
\]

Next, we have that 
\[
\left | \sum_{(x_1, \dots x_k) \in \mathcal{X}^k} C((x_1, \dots x_k)) \cdot \left [\widetilde{X}_{0}^{\otimes k}(x_1, \dots x_k) - \widetilde{X}_{1}^{\otimes k}(x_1, \dots x_k) \right ] \right |
\]
\[
\geq \left | \sum_{(x_1, \dots x_k) \in \mathcal{X}^k} C((x_1, \dots x_k)) \cdot \left [\widetilde{X}_{0, \delta}^{\otimes k}(x_1, \dots x_k) - \widetilde{X}_{1, \delta}^{\otimes k}(x_1, \dots x_k) \right ] \right | 
\]
\[
- \left | \sum_{(x_1, \dots x_k) \in \mathcal{X}^k} C((x_1, \dots x_k)) \cdot \left [\widetilde{X}_{0}^{\otimes k}(x_1, \dots x_k) - \widetilde{X}_{0, \delta}^{\otimes k}(x_1, \dots x_k) \right ] \right | 
\]
\[
- \left | \sum_{(x_1, \dots x_k) \in \mathcal{X}^k} C((x_1, \dots x_k)) \cdot \left [\widetilde{X}_{1}^{\otimes k}(x_1, \dots x_k) - \widetilde{X}_{1, \delta}^{\otimes k}(x_1, \dots x_k) \right ] \right | 
\]
\[
\geq \left | \sum_{(x_1, \dots x_k) \in \mathcal{X}^k} C((x_1, \dots x_k)) \cdot \left [\widetilde{X}_{0, \delta}^{\otimes k}(x_1, \dots x_k) - \widetilde{X}_{1, \delta}^{\otimes k}(x_1, \dots x_k) \right ] \right |  
\]
\[
- d_{\mathrm{TV}}(\widetilde{X}_{0}^{\otimes k}, \widetilde{X}_{0, \delta}^{\otimes k}) - d_{\mathrm{TV}}(\widetilde{X}_{1}^{\otimes k}, \widetilde{X}_{1, \delta}^{\otimes k})
\]
\[
\geq d_{\mathrm{TV}}(\widetilde{X}_{0}^{\otimes k}, \widetilde{X}_{1}^{\otimes k}) - 4\delta k.
\]

Finally, observe that because our circuit $C$ only distinguishes samples $x$ based on \emph{the partition} that the sample is in, $C$ achieves the same distinguishing advantage on $X_0$ vs. $\widetilde{X}_0$ (and likewise for $X_1$) because of Item (b) of \cref{thm:MCproperties_restatement}. That is to say, 
\[
\left | \sum_{(x_1, \dots x_k) \in \mathcal{X}^k} C((x_1, \dots x_k)) \cdot \left [\widetilde{X}_{0}^{\otimes k}(x_1, \dots x_k) - \widetilde{X}_{1}^{\otimes k}(x_1, \dots x_k) \right ] \right |
\]\[
= \left | \sum_{(x_1, \dots x_k) \in \mathcal{X}^k} C((x_1, \dots x_k)) \cdot \left [X_{0}^{\otimes k}(x_1, \dots x_k) - X_{1}^{\otimes k}(x_1, \dots x_k) \right ] \right |.
\]

Hence, there is a circuit in \[\mathcal{F}_{qk, tk + O(rk) + O(k \cdot \polylog(r / \delta))},
\]
which distinguishes $X_0^{\otimes k}, X_1^{\otimes k}$ with advantage $d_{\mathrm{TV}}(\widetilde{X}_{0}^{\otimes k}, \widetilde{X}_{1}^{\otimes k}) - 4\delta k$.
\end{proof}

\section{Defining One Intermediate Variable}\label{section:one-intermediate}

Instead of relating the distinguishability of $X_0, X_1$ to the total variation distance between \textit{two} intermediate random variables $\widetilde{X}_0, \widetilde{X}_1$, one may instead be interested in if similar results can be obtained with only \textit{one} intermediate random variable, i.e., setting $\widetilde{X}_1 = X_1$. As stated in \Cref{thm:main} Part (4), similar results hold regarding indistinguishability and distinguishability, with some modifications to the circuit size in the distinguishability component of this theorem. An important special case is when $X_1$ is the uniform distribution over $\mathcal{X}$. 

\begin{definition}\label{def:hat_variable}
    For a pair of random variables $X_0, X_1$, class $\mathcal{F}$ of functions, and parameter $\eps$. We define parameter $\eps'$, the family of functions $\mathcal{F}'$, distribution $\mathcal{D}$, function $g$, random variable $\widehat{X}_0$, and function $p: \mathcal{X} \to [m]$ as follows.
    \begin{enumerate}[label=(\alph*)]
        \item Let $\eps' = \eps^2$.
        \item Let $\mathcal{F}'$ be the family of functions such that $\mathcal{F}_{c \log |\mathcal{X}|, c} = \mathcal{F}'$, for some universal constant $c$.
        \item Let the distribution $\mathcal{D}$ be as follows: To sample from $\mathcal{D}$, first pick $B \sim \{0, 1\}$ uniformly at random. Output a sample $x \sim X_B$. Note that $$\mathbb{P}_{x \sim \mathcal{D}}[x] = \frac{1}{2} \mathbb{P}[X_0 = x] + \frac{1}{2}\mathbb{P}[X_1 = x].$$

        For a subset of the domain $\mathcal{S} \subseteq \mathcal{X}$, let $\mathcal{D}_{| \mathcal{S}}$ be the conditional distribution of $\mathcal{D}$ over the set $\mathcal{S}$.
        \item We then define the randomized function $g$ as:
        $$g(x) = \begin{cases}
            0 ~~ \text{with probability } \frac{\mathbb{P}[X_0 = x]}{\mathbb{P}[X_0 = x] + \mathbb{P}[X_1 = x]} \\
            1 ~~ \text{with probability } \frac{\mathbb{P}[X_1 = x]}{\mathbb{P}[X_0 = x] + \mathbb{P}[X_1 = x]}.
        \end{cases}$$
        \item Random variable $\widehat{X}_0$: Consider the multicalibrated partition $\mathcal{P} = \{P_i\}$ guaranteed by the Multicalibration Theorem when applied to the function $g$, distribution $\mathcal{D}$ over domain $\mathcal{X}$, class of functions $\mathcal{F}'$, and parameters $\eps'$ and $\gamma = (\eps')^2$ to obtain a $(\mathcal{F}', \eps', \gamma)$-approximately multicalibrated partition $\mathcal{P}$ of $\mathcal{X}$. 
        Given the multicalibrated partition, construct the random variable $\widehat{X}_0$ as follows: to sample according to $\widehat{X}_0$, first choose a piece of the partition $P_i \in \mathcal{P}$, where $P_i$ is chosen with probability $\mathbb{P}[X_0 \in P_i]$. If $\alpha_1(P_i) \geq \sqrt{\eps'}$, then sample $x \in P_i$ according to $X_1 |_{P_i}$. Otherwise, sample $x$ according to $X_{0}|_{P_i}$. \textit{This final conditioning step is the key difference from \cref{def:tilde_variables}.}
        \item We let $p: \mathcal{X} \to [m]$ be the function that returns which part of the multicalibrated partition $\mathcal{P}$ an element $x \in \mathcal{X}$ is in. That is, if $x \in P_i$ for $P_i \in \mathcal{P}$, $p(x) = i$.
    \end{enumerate}
\end{definition} 

We prove the following theorem, which is a general version of \Cref{thm:main} Part (4). 

\begin{theorem}\label{thm:oneIntermediateMain}
    For every pair of random variables $X_0, X_1$ over $\mathcal{X}$, any family of functions $\mathcal{F}$ and every $\eps > 0$, there exists a random variable $\widehat{X}_0$ such that for every $k>0$,
    \begin{enumerate}
        \item $\widehat{X}_0$ is $\eps$-indistinguishable from $X_0$ by functions in $\mathcal{F}$.
        \item  $X_0^{\otimes k}$ and $X_1^{\otimes k}$ are $(d_{TV}((\widehat{X}_0)^{\otimes k}, X_1^{\otimes k})+ k \cdot \eps)$-indistinguishable by functions in $\mathcal{F}^{\otimes k}$.
        \item $X_0^{\otimes k}$ and $X_1^{\otimes k}$ are $(d_{\mathrm{TV}}((\widehat{X}_0)^{\otimes k}, X_1^{\otimes k}) - 3k \eps)$-distinguishable by functions in \[
     \mathcal{F}^{\otimes k}_{K_2, K_1}
    \]
    with $K_1 = O(k \cdot \log(|\mathcal{X}| / \eps) \cdot \log(|\mathcal{X}|)/ \eps^{24} ) + O(k \log(|\mathcal{X}|) \cdot \mathrm{polylog}(1 / \eps))$, and $K_2 = O(k / \eps^{12})$.
    \end{enumerate}
\end{theorem}

\begin{proof}
    We take $\widehat{X}_0$ to be defined as in \cref{def:hat_variable}. Item (1) of the theorem follows from \cref{lem:item1ofthm52}, Item (2) follows from \cref{thm:computational-indistinguishability-3}, and Item (3) follows from \cref{lem:item3thm52} where we set $\delta = \eps / 4$.
\end{proof}

\subsection{Computational Indistinguishability}
We prove that $\widehat{X}_0$ is $O(\eps)$-indistinguishable from $X_0$ by circuits of size $s$. Let $H_{1/2}(Y)$ be the R\'enyi entropy of $Y$ of order $1/2$, as in \Cref{renyi}.

\begin{lemma}[Item (2) of \cref{thm:oneIntermediateMain}]\label{thm:computational-indistinguishability-3}
    For every pair of random variables $X_0, X_1$, every family of functions $\mathcal{F}$, and every $\eps>0$, and for $\widehat{X}_0$ as defined in \cref{def:hat_variable}, then $X_0^{\otimes k}$ and $X_1^{\otimes k}$ are $(d_{TV}((\widehat{X}_0)^{\otimes k}, X_1^{\otimes k})+ k \cdot \eps)$-indistinguishable by functions in $\mathcal{F}^{\otimes k}$.
\end{lemma}

Towards proving \Cref{thm:computational-indistinguishability-3}, we begin by proving the following lemma.

\begin{lemma}[Item (1) of \cref{thm:oneIntermediateMain}]\label{lem:item1ofthm52}
    For $X_0, X_1, \mathcal{F}, \eps > 0$, define $\widehat{X}_0$ as in \cref{def:hat_variable}. Then, $X_0$ is $(\mathcal{F}, O(\eps))$-indistinguishable from $\widehat{X}_0$.
\end{lemma}

\begin{proof}
Define three sets partitioning the set of $P_i \in \mathcal{P}$. First, let $\mathcal{P}_1$ be the set of partitions such that $\mathbb{P}_{x \sim \mathcal{D}}[x \in P_i] \geq \gamma$ and $\alpha_1(P_i) \geq \sqrt{\eps'} = \eps$. Let $\mathcal{P}_2$ be the set of partitions such that $\mathcal{P}_{x \sim \mathcal{D}}[x \in P_i] \geq \gamma$ and $\alpha_1(P_i) < \sqrt{\eps'} = \eps$. Let $\mathcal{P}_3$ be the remaining partitions, which satisfy $\mathbb{P}_{x \sim \mathcal{D}}[x \in P_i] < \gamma$.

First, observe that: 
$$\left|\sum_{P_i \in \mathcal{P}} \sum_{x \in P_i} f(x) \cdot \left( \mathbb{P}[X_0 = x] -  \mathbb{P}[\widehat{X}_0 = x]\right) \right| \leq \sum_{j = 1}^3 \sum_{P_i \in \mathcal{P}_j} \left| \sum_{x \in P_i} f(x) \cdot \left( \mathbb{P}[X_0 = x] -  \mathbb{P}[\widehat{X}_0 = x]\right) \right|.$$

Let us focus on the first on the summation over $P_i \in \mathcal{P}_1$. Observe that on all $P_i \in \mathcal{P}_1$, by definition $\widehat{X}^{(\mathcal{U})}_{0}|_{P_i} = \mathcal{U}|_{P_i}$, the uniform distribution over elements of $P_i$. This gives the following:
$$\sum_{P_i \in \mathcal{P}_1} \left| \sum_{x \in P_i} f(x) \cdot \left( \mathbb{P}[X_0 = x] -  \mathbb{P}[\widehat{X}_0 = x]\right) \right|$$
$$= \sum_{P_i \in \mathcal{P}_1} \mathbb{P}[X_0 \in P_i]  \left| \sum_{x \in P_i} f(x) \cdot \left( \mathbb{P}[X_{0}|_{P_i} = x] -  \mathbb{P}[\mathcal{U}|_{P_i} = x]\right) \right|$$
$$= \sum_{P_i \in \mathcal{P}_1} \alpha_0(P_i) \cdot 2 \mathbb{P}[\mathcal{D} \in P_i]  \left| \sum_{x \in P_i} f(x) \cdot \left( \mathbb{P}[X_{0}|_{P_i} = x] -  \mathbb{P}[\mathcal{U}|_{P_i} = x]\right) \right|$$
Using Lemma \ref{lem:X0Pi_vs_X1Pi} to bound the indistinguishability of $X_{0}|_{P_i}$ and $\mathcal{U}|_{P_i}$, we obtain:
$$\leq \sum_{P_i \in \mathcal{P}_1}  \frac{\eps' \cdot \alpha_0(P_i) \cdot 2 \mathbb{P}[\mathcal{D} \in P_i]}{ \alpha_0(P_i) \alpha_1(P_i)} \leq  \sum_{P_i \in \mathcal{P}_1}  \frac{\eps' \cdot 2 \mathbb{P}[\mathcal{D} \in P_i]}{\alpha_1(P_i)}.$$
Because $\alpha_1(P_i) \geq \sqrt{\eps'}$ for all $P_i \in \mathcal{P}$, this is:
$$\leq 2 \sqrt{\eps'} \sum_{P_i \in \mathcal{P}_1} \mathbb{P}[\mathcal{D} \in P_i] \leq 2 \sqrt{\eps'} = 2 \eps.$$

We now shift to bounding the summation over $P_i \in \mathcal{P}_2$. All $P_i \in \mathcal{P}_2$ satisfy $\mathcal{P}_{x \sim \mathcal{D}}[x \in P_i] \geq \gamma$ and $\alpha_1(P_i) < \sqrt{\eps'}$. By definition of $\widehat{X}_0$, for all $P_i \in \mathcal{P}_2$ and all $x \in P_i$, $\mathbb{P}[\widehat{X}_0 = x] = \mathbb{P}[X_{0} = x]$.
This implies that 
$$\sum_{P_i \in \mathcal{P}_2} \left| \sum_{x \in P_i} f(x) \cdot \left( \mathbb{P}[X_0 = x] -  \mathbb{P}[\widehat{X}_0 = x]\right) \right| = 0.$$

Lastly, we look at the sum over $P_i \in \mathcal{P}_3$, for which $\mathbb{P}_{x \sim \mathcal{D}}[x \in P_i] < \gamma$. In this case, exactly as in the proof of \Cref{thm:MCproperties} Part (a),
$$\sum_{P_i \in \mathcal{P}_3} \left|\sum_{x \in P_i} f(x) \cdot \left( \mathbb{P}[X_0 = x] -  \mathbb{P}[\widehat{X}_0 = x]\right) \right| < 4 \gamma |\mathcal{P}|.$$
By definition of $\gamma$ and the bound on $|\mathcal{P}|$ from multicalibration, $4 \gamma |\mathcal{P}|$ is $O(\eps')$, which is $O(\eps^2)$.

Putting all of the cases together, we can conclude that 
$$\left|\sum_{P_i \in \mathcal{P}} \sum_{x \in P_i} f(x) \cdot \left( \mathbb{P}[X_0 = x] -  \mathbb{P}[\widehat{X}_0 = x]\right) \right| \leq O(\sqrt{\eps'}) = O(\eps).$$ Therefore, $X_0$ is $(\mathcal{F}, O(\eps))$-indistinguishable from $\widehat{X}_0$.
\end{proof}

The indistinguishability can be extended to product distributions when we make an additional assumption about the family $\mathcal{F}$ being closed under marginals.

\begin{lemma}\label{lem:moving-to-product-distributions-uniform}
    For $X_0, X_1, \mathcal{F}, \eps > 0$, define $\widehat{X}_0$ as in \cref{def:hat_variable}. Then, for $f \in \mathcal{F}^{\otimes k}$,
    $$\left| \sum_{\Bar{x} \in \mathcal{X}^{k}} f(\Bar{x}) \cdot \left(\mathbb{P}[X_{0}^{\otimes k} = \Bar{x}] - \mathbb{P}[(\widehat{X}_0)^{\otimes k} = \Bar{x}] \right) \right| \leq O(k \cdot \eps).$$
\end{lemma}

\begin{proof}
    This is exactly the same as the proof of \Cref{lem:moving-to-product-distributions}.
\end{proof}

\begin{lemma}\label{lemma:computational-indistinguishability-bound-2}[\Cref{thm:computational-indistinguishability-3}]
    For $X_0, X_1, \mathcal{F}, \eps > 0$, define $\widehat{X}_0$ as in \cref{def:hat_variable}. For $f \in \mathcal{F}^{\otimes k}$, the following three inequalities hold:
$$\left| \sum_{\Bar{x} \in \mathcal{X}^{k}} f(\Bar{x}) \cdot \left(\mathbb{P}[X_{0}^{\otimes k} = \Bar{x}] - \mathbb{P}[X_1^{\otimes k} = \Bar{x}] \right) \right| \leq O(k \cdot \eps) + d_{\mathrm{TV}}\left((\widehat{X}_0)^{\otimes k}, X_1^{\otimes k}\right)$$ $$\leq O(k \cdot \eps) + \sqrt{2} \sqrt{1 - \left(1 - d_H^2\left(\widehat{X}_0, X_1\right)\right)^k}$$ $$\leq O(k \cdot \eps)  + \sqrt{2k} \cdot d_H\left(\widehat{X}_0, X_1\right).$$
\end{lemma}

\begin{proof}
    The proof follows very similarly to the proof of \Cref{lemma:computational-indistinguishability-bound}, except that in the triangle inequality below we break our term into two pieces instead of three because we only introduce a variable $\widehat{X}_0$ for $X_0$ and not $X_1$.

     For $f \in \mathcal{F}^{\otimes k}$,
$$\left| \sum_{\Bar{x} \in \mathcal{X}^{k}} f(\Bar{x}) \cdot \left(\mathbb{P}[X_{0}^{\otimes k} = x] - \mathbb{P}[X_1^{\otimes k} = x] \right) \right| \leq \left| \sum_{\Bar{x} \in \mathcal{X}^{k}} f(\Bar{x}) \cdot \left(\mathbb{P}[X_{0}^{\otimes k} = \Bar{x}] - \mathbb{P}[(\widehat{X}_0)^{\otimes k} = \Bar{x}] \right) \right|$$
$$+ \left| \sum_{\Bar{x} \in \mathcal{X}^{k}} f(\Bar{x}) \cdot \left(\mathbb{P}\left[(\widehat{X}_0)^{\otimes k} = \Bar{x}\right] - \mathbb{P}\left[X_1^{\otimes k} = \Bar{x}\right] \right) \right|$$
$$\leq O(k \cdot \eps) + d_{\mathrm{TV}}\left((\widehat{X}_0)^{\otimes k}, X_1^{\otimes k}\right)$$
$$\leq O(k \cdot \eps) + \sqrt{2} \sqrt{1 - \left( 1 - d_H^2\left(\widehat{X}_0, X_1\right)\right)^k}$$
$$\leq  O(k \cdot \eps)  + \sqrt{2k} \cdot d_H\left(\widehat{X}_0, X_1\right). \eqno \qedhere$$
\end{proof}

\subsection{Moving from Total Variation Distance to Partitions}

Before moving to efficient distinguishers, we prove the following claim regarding $\widehat{X}_0$. This will allow us to construct efficient distinguishers based on the distributions over partitions induced by $\widehat{X}_0$ and $X_1$, as done before for the case of distinguishing $X_0$ versus general $X_1$.

\begin{claim}\label{clm:uniformPartitionTV}
    For $X_0, X_1, \mathcal{F}, \eps > 0$, define $\widehat{X}_0$ and $\mathcal{P}$ as in \cref{def:hat_variable}. We have that:
    \[
    d_{\mathrm{TV}}\left(p(\widehat{X}_0), p\left(X_1\right)\right) \leq d_{\mathrm{TV}}\left(\widehat{X}_0, X_1\right) \leq d_{\mathrm{TV}}\left(p(\widehat{X}_0), p\left(X_1\right)\right) + \eps.
    \]
\end{claim}

\begin{proof}
    For the first inequality, observe that in general, for any random variables $X, Y$, and any function $f$ on their domain, it is the case that $d_{\mathrm{TV}}(f(X), f(Y)) \leq d_{\mathrm{TV}}(X, Y)$.

    For the second inequality, observe that we can express the total variation distance as the following, where $\eps' := \eps^2$ as per \cref{def:hat_variable}:
    \[
    d_{\mathrm{TV}}\left(\widehat{X}_0, X_1\right) = 
    \sum_{\substack{x \in \mathcal{X} \\ X_1(x) > \widehat{X}_0(x)}} \left(X_1(x) - \widehat{X}_0(x)\right)
    \]
    $$=\sum_{\substack{P_i \in \mathcal{P} \\ \alpha_1(P_i) \geq \eps}} \sum_{\substack{x \in P_i \\ X_1(x) > \widehat{X}_0(x)}} \left(X_1(x) - \widehat{X}_0(x)\right) + \sum_{\substack{P_i \in \mathcal{P} \\ \alpha_1(P_i) < \eps}} \sum_{\substack{x \in P_i \\ X_1(x) > \widehat{X}_0(x)}} \left(X_1(x) - \widehat{X}_0(x)\right).$$
    Let us now bound each of the two terms (i.e., each double summation).
    
    First, let us bound the summation over $P_i \in \mathcal{P}$ such that $\alpha_1(P_i) \geq \eps = \sqrt{\eps'}$. For each such $P_i$, the distributions $X_1|_{P_i}$ and $\widehat{X}_{0}|_{P_i}$ are exactly the same. Thus, the ability to distinguish $X_1$ and $\widehat{X}_0$ in these parts is exactly determined by the probability of each partition appearing, which is bounded by $d_{\mathrm{TV}}(p(\widehat{X}_0), p(X_1))$.

    Second, we bound the summation over $P_i \in \mathcal{P}$ such that $\alpha_1(P_i) < \eps = \sqrt{\eps'}$. Recall that if $P_i$ satisfies $\alpha_1(P_i) < \sqrt{\eps'}$, this means that $$\frac{\Pr[X_1 \in P_i]}{\Pr[X_1 \in P_i] + \Pr[\widehat{X}_0 \in P_i]} < \sqrt{\eps'}.$$ Therefore, in particular, for sufficiently small, constant $\eps$, we have 
        \[
        \frac{\Pr[X_1 \in P_i]}{\Pr[\widehat{X}_0 \in P_i]} < 2 \sqrt{\eps'} = 2 \eps.
        \]
        Now, we observe that
        \[
        \sum_{\substack{P_i \in \mathcal{P} \\ \alpha_i(P_i) < \eps}} \sum_{\substack{x \in P_i \\ X_1(x) > \widehat{X}_0(x)}} \left(X_1(x) - \widehat{X}_0(x)\right) \leq \sum_{\substack{P_i \in \mathcal{P} \\ \alpha_i(P_i) < \eps}} \sum_{\substack{x \in P_i \\ X_1(x) > \widehat{X}_0(x)}} X_1(x) 
        \]
        \[
        \leq \sum_{\substack{P_i \in \mathcal{P} \\ \alpha_i(P_i) < \eps}} \Pr[X_1 \in P_i]  < \sum_{\substack{P_i \in \mathcal{P} \\ \alpha_i(P_i) < \eps}} 2 \eps \Pr[\widehat{X}_0 \in P_i] \leq 2 \eps,
        \]
        where in the final inequality, we used the fact that $\sum_{P_i \in \mathcal{P}}\Pr[\widehat{X}_0 \in P_i] = 1$.

    Together, the bounds on the two double summations imply the following bound:
    $$
    d_{\mathrm{TV}}(\widehat{X}_0, X_1) \leq d_{\mathrm{TV}}\left(p(\widehat{X}_0), p\left(X_1\right)\right) + 2 \eps. \eqno\qedhere
    $$
\end{proof}

In fact however, we will need a version of the above claim which extends to sampling $k$ times. We present this version of the claim below, which uses much of the same logic:

\begin{claim}\label{clm:uniformPartitionTVkSamples}
    For $X_0, X_1, \mathcal{F}, \eps > 0$, define $\widehat{X}_0$ and $\mathcal{P}$ as in \cref{def:hat_variable}. For an integer $k$, and for a sufficiently small $\eps$, we have that:
    \[
    d_{\mathrm{TV}}\left(p\left(\widehat{X}_0\right)^{\otimes k}, p\left(X_1\right)^{\otimes k}\right) \leq d_{\mathrm{TV}}\left(\left (\widehat{X}_0 \right )^{\otimes k}, \left ( X_1\right) \right )^{\otimes k} \leq d_{\mathrm{TV}}\left(p\left(\widehat{X}_0\right)^{\otimes k}, p\left(X_1\right)^{\otimes k}\right) + 2k \eps.
    \]
\end{claim}

\begin{proof}
    For the first inequality, observe that in general, for any random variables $X, Y$, and any function $f$ on their domain, it is the case that $d_{\mathrm{TV}}(f(X), f(Y)) \leq d_{\mathrm{TV}}(X, Y)$. This still holds under taking $k$-samples.

    For the second inequality, observe that we can express the total variation distance as:
    \[
    d_{\mathrm{TV}}\left(\left (\widehat{X}_0 \right )^{\otimes k}, \left ( X_1\right) \right )^{\otimes k}= 
    \sum_{\substack{(x_1, \dots x_k) \in \mathcal{X}^{k} \\ X_1^{\otimes k}(x_1, \dots x_k) > \left (\widehat{X}_0\right )^{\otimes k}(x_1, \dots x_k)}} \left(X_1^{\otimes k}(x_1, \dots x_k) - \left(\widehat{X}_0\right )^{\otimes k}(x_1, \dots x_k) \right)
    \]
    \[
    =\sum_{\substack{P^{(1)}, \dots P^{(k)} \in \mathcal{P}^{\otimes k} \\ \alpha_1(P^{(i)}) \geq \eps}} \sum_{\substack{(x_1, \dots x_k): x_i \in P^{(i)} \\ X_1^{\otimes k}(x_1, \dots x_k) > \left (\widehat{X}_0\right )^{\otimes k}(x_1, \dots x_k)}} \left(X_1^{\otimes k}(x_1, \dots x_k) - \left(\widehat{X}_0\right )^{\otimes k}(x_1, \dots x_k)\right)
    \]
    \[
    + \sum_{\substack{P^{(1)}, \dots P^{(k)} \in \mathcal{P}^{\otimes k} \\ \exists i: \alpha_1(P^{(i)}) < \eps}} \sum_{\substack{(x_1, \dots x_k): x_i \in P^{(i)} \\ X_1^{\otimes k}(x_1, \dots x_k) > \left (\widehat{X}_0\right )^{\otimes k}(x_1, \dots x_k)}} \left(X_1^{\otimes k}(x_1, \dots x_k) - \left(\widehat{X}_0\right )^{\otimes k}(x_1, \dots x_k)\right).\]
    Let us now bound each of the two terms (i.e., each double summation).
    
    We start by bounding the first term, i.e., the sum over pieces $P^{(1)}, \dots P^{(k)}$ such that $\alpha_0({P^{(i)}}) \geq \eps = \sqrt{\eps'}$. Observe that, by definition, on these pieces the marginal distributions $X_1|_{P_i}$ and $\widetilde{X}^{(X_1)}_{0}|_{P_i}$ are exactly the same. Thus, the ability to distinguish $X_1$ and $\widehat{X}_0$ in these parts is exactly determined by the probability of each partition appearing. Thus, we can bound this term by $d_{\mathrm{TV}}\left(p\left(\widehat{X}_0\right)^{\otimes k}, p\left(X_1\right)^{\otimes k}\right)$.

    Next, we bound the summation over terms $x_1, \dots x_k$ such that there exist an $i$ for which $x_i \in P^{(i)}$ and $P^{(i)} \in \mathcal{P}$ such that $\alpha_1(P^{(i)} ) < \eps = \sqrt{\eps'}$ (this equality comes as per \cref{def:hat_variable}). Recall that if $P^{(i)} $ satisfies $\alpha_1(P^{(i)} ) < \sqrt{\eps'}$, this means that $$\frac{\Pr[X_1 \in P^{(i)} ]}{\Pr[X_1 \in P^{(i)} ] + \Pr[\widehat{X}_0 \in P^{(i)} ]} < \sqrt{\eps'}.$$ Therefore, in particular, for sufficiently small, constant $\eps$, we have 
        \[
        \frac{\Pr[X_1 \in P^{(i)} ]}{\Pr[\widehat{X}_0 \in P^{(i)} ]} < 2 \sqrt{\eps'} = 2 \eps.
        \]
        Now, we observe that
        \[
        \sum_{\substack{P^{(1)}, \dots P^{(k)} \in \mathcal{P}^{\otimes k} \\ \exists i: \alpha_1(P^{(i)}) < \eps}} \sum_{\substack{(x_1, \dots x_k): x_i \in P^{(i)} \\ X_1^{\otimes k}(x_1, \dots x_k) > \left (\widehat{X}_0\right )^{\otimes k}(x_1, \dots x_k)}} \left(X_1^{\otimes k}(x_1, \dots x_k) - \left(\widehat{X}_0\right )^{\otimes k}(x_1, \dots x_k)\right)\]
        \[
\leq \sum_{\substack{P^{(1)}, \dots P^{(k)} \in \mathcal{P}^{\otimes k} \\ \exists i: \alpha_1(P^{(i)}) < \eps}} \sum_{\substack{(x_1, \dots x_k): x_i \in P^{(i)} \\ X_1^{\otimes k}(x_1, \dots x_k) > \left (\widehat{X}_0\right )^{\otimes k}(x_1, \dots x_k)}} \left(X_1^{\otimes k}(x_1, \dots x_k)\right)
        \]
        \[
        \leq \sum_{i = 1}^k \sum_{\substack{P^{(i)} \in \mathcal{P} \\ \alpha_0(P^{(i)}) < \sqrt{\eps}}} \Pr[X_1 \in P_i]  < \sum_{i = 1}^k \sum_{\substack{P^{(i)} \in \mathcal{P} \\ \alpha_0(P^{(i)}) < \eps}} 2 \eps \Pr[\widehat{X}_0 \in P_i] \leq 2k \eps,
        \]
        where in the final inequality, we used the fact that $\sum_{P_i \in \mathcal{P}}\Pr[\widehat{X}_0 \in P_i] = 1$.

    Together, the bounds on the two double summations imply the following bound:
    $$
    d_{\mathrm{TV}}\left(\left (\widehat{X}_0 \right )^{\otimes k}, \left ( X_1\right)^{\otimes k} \right ) \leq d_{\mathrm{TV}}\left(p\left(\widehat{X}_0\right)^{\otimes k}, p\left(X_1\right)^{\otimes k}\right) + 2k \eps. \eqno\qedhere
    $$
\end{proof}

\subsection{Efficient Distinguishers}

In this section, we show the following:

\begin{lemma}[[Item (3) of \cref{thm:oneIntermediateMain}]\label{lem:item3thm52}
    For $X_0, X_1, \mathcal{F}, \eps > 0$, define $\widehat{X}_0$ and $\mathcal{P}$ as in \cref{def:hat_variable}. Then, there exists a circuit in 
    \[
     \mathcal{F}_{K_2, K_1}
    \]
    with $K_1 = O(k / \eps^{24} \cdot \log(|\mathcal{X}|)\cdot \log(|\mathcal{X}| / \eps)) + O(k \mathrm{polylog}(1 / \eps \delta))$, and $K_2 = O(k / \eps^{12})$, which distinguishes $X^{\otimes k}_0, X_1^{\otimes k}$ with advantage $d_{\mathrm{TV}}((\widehat{X}_0)^{\otimes k}, X_1^{\otimes k}) - 4\delta k - 2k \eps$.
\end{lemma}

\begin{proof}
    We follow the same proof as in \cref{thm:distinguishingCircuit}. First, recall that because the Multicalibration Theorem we use to construct the partition $\mathcal{P} = \{P_i\}$ is efficient, we can construct small circuits that compute the identity of the partition $P_i: x \in P_i$ for any $x \in \mathcal{X}$. Further, the number of parts in the partition is sufficiently small such that we can implement the optimal distinguisher over these parts.  
    
    Specifically, as in \cref{thm:distinguishingCircuit}, we build the optimal distinguisher over the partition random variables $p(\widehat{X}_0)^{\otimes k},  p(X_1)^{\otimes k}$ (up to some additive error of $\delta$), with circuits in 
    \[
    \mathcal{F}_{qk, tk\log(|\mathcal{X}|) + O(rk) + O(k \cdot \polylog(r / \delta))},
    \]
    where $q = O(1 / \eps'^6), r = O(1 / \eps'), t = O(1 /(\eps'^4 \gamma) \cdot \log(|\mathcal{X}|)/\eps' )$. By plugging in $\eps' = \eps^2$, we see that this circuit lies in 
    \[
    \mathcal{F}_{K_2, K_1},
    \]
    where $K_1 = O(k / \eps^{24} \cdot \log(|\mathcal{X}|) \cdot \log(|\mathcal{X}| / \eps)) + O(k \mathrm{polylog}(1 / \eps \delta))$, and $K_2 = O(k / \eps^{12})$.

    We denote this circuit by $C$, and formally, $C$ achieves: 
    \[
\left | \sum_{(x_1, \dots x_k) \in \mathcal{X}^k} C((x_1, \dots x_k)) \cdot \left [(\widehat{X}_0)^{\otimes k}, X_1^{\otimes k})(x_1, \dots x_k) - X_1^{\otimes k}(x_1, \dots x_k) \right ] \right |
\]
\[
\geq d_{\mathrm{TV}}(p((\widehat{X}_0)^{\otimes k}), p(X_1^{\otimes k})) - 4 \delta k\geq d_{\mathrm{TV}}((\widehat{X}_0)^{\otimes k}, X_1^{\otimes k}) - 4\delta k - 2 k \eps,
\]
where we used \cref{clm:uniformPartitionTVkSamples} in the final step. 

Finally, we remark that because the circuit $\mathcal{C}$ depends only on the labels of the partitions that each element is in, $\mathcal{C}$ distinguishes $X_0^{\otimes k}, X_1^{\otimes k}$ with the same advantage that it distinguishes $(\widehat{X}_0)^{\otimes k}, X_1^{\otimes k}$, as $X_0^{\otimes k}$ and $(\widehat{X}_0)^{\otimes k}$ place the same mass on each partition by definition. This then yields the claim that there is a circuit in $\mathcal{F}_{qk, tk\log(|\mathcal{X}|) + O(rk) + O(k \cdot \polylog(r / \delta))}$ which distinguishes $X^{\otimes k}_0, X_1^{\otimes k}$ with advantage $d_{\mathrm{TV}}((\widehat{X}_0)^{\otimes k}, X_1^{\otimes k}) - 4\delta k- 2 k \eps $.
\end{proof}

Note that this claim then immediately implies Item (3) of \cref{thm:oneIntermediateMain} by setting $\delta = \eps / 4$.

\section{Re-deriving a Result of Halevi and Rabin}\label{section:geier}

We now show our results and analysis imply a result comparable to the main theorem proven in \cite{HR08} and \cite{geier2022tight}. These papers prove a version of the following bound on the statistical distance of product distributions that instead holds with respect to computational indistinguishability:
\begin{equation}\label{eq:TVD-product}
    d_{\mathrm{TV}}(X^{\otimes k}, Y^{\otimes k}) \leq 1 - (1 - d_{\mathrm{TV}}(X, Y))^{k}.
\end{equation}
Geier's result has slightly improved parameters and establishes the following theorem:

\begin{theorem}[\cite{geier2022tight}]\label{thm:geier}
    Let $X_0$ and $X_1$ be distributions over $b$ bits that are $d$-indistinguishable for circuits of size $s$. For every $k \in \mathbb{N}$ and $\delta > 0$, $X_0^{\otimes k}$ and $X_1^{\otimes k}$ are $(1 - (1 - d)^k + \delta)$-indistinguishable for circuits of size $s_{k, \delta}$, defined as follows:
   $$s_{k, \delta} := \frac{s-1}{c_{k, \delta}} - 5 k b - 1 ~~ \text{ where } ~~ c_{k, \delta} := \left\lceil \frac{\log(b \delta)}{\log(1 - (1 - d)^k + \delta)}\right\rceil.$$
\end{theorem}

The approach Geier uses to prove this theorem differs from that used in the statistical distance setting to prove Equation (\ref{eq:TVD-product}). On the other hand, using the results we've proven through a multicalibration-based approach, we can imply a similar result \textit{through statistical distance} by relating the computational indistinguishability of $X_0$ and $X_1$ and $X_0^{\otimes k}$ and $X_1^{\otimes k}$ to the statistical distances between the random variables $\widetilde{X}_0$ and $\widetilde{X}_1$. We prove that the following theorem as an immediate implication of the results proven in previous sections.

As in previous sections, $\mathcal{F}'$ be the family of functions computable by circuits of size at most $s \cdot c + c \cdot \log |\mathcal{X}|$ for a universal constant $c$. Also as in previous sections, let $\mathcal{F}$ be the family of functions computable by circuits of size at most $s$. Define the family $\mathcal{F}''$ to be (for a parameter $\eps$):
    \[ \mathcal{F}'' :=
    \mathcal{F}'_{K_1, K_2},
    \]
    with $K_1 = O(\frac{1}{\eps^{12}} \cdot \log(|\mathcal{X}|)/\eps ) + O(\polylog(1 /\eps))$, and $K_2 = O(1/\eps^6)$.
\begin{theorem}\label{thm:geier-implication} Let $X_0$ and $X_1$ be distributions over domain $\mathcal{X}$ that are $d$-indistinguishable for functions in $\mathcal{F}''$. For every $k \in \mathbb{N}$, there is a very small $\delta = O(k \epsilon)$ such that $X_0^{\otimes k}$ and $X_1^{\otimes k}$ are $(1 - (1 - d)^k + \delta )$-indistinguishable for functions in $\mathcal{F}$.
\end{theorem}

\begin{proof}
From Lemma \ref{lemma:computational-indistinguishability-bound}, for all $f \in \mathcal{F}$, $$\left| \sum_{\Bar{x} \in \mathcal{X}^{k}} f(\Bar{x}) \cdot \left(\mathbb{P}[X_{0}^{\otimes k} = \Bar{x}] - \mathbb{P}[X_{1}^{\otimes k} = \Bar{x}] \right) \right| \leq O(k \eps) + d_{\mathrm{TV}}(\widetilde{X}_0^{\otimes k}, \widetilde{X}_1^{\otimes k}).$$
For the random variables $X_0, X_1$, family $\mathcal{F}$ of functions, and parameter $\eps \leq \frac{1}{100k}$ (or $\eps = \frac{1}{C k}$ for any sufficiently large constant $C$), consider the construction of random variables $\widetilde{X}_0, \widetilde{X}_1$ as promised in \cref{thm:MCproperties_restatement}. By Equation \ref{eq:TVD-product} and Lemma \ref{thm:distinguishingCircuit}, the above is
$$\leq O(k \eps) + 1 - (1 - d_{\mathrm{TV}}(\widetilde{X}_0, \widetilde{X}_1))^k $$
$$\leq O(k \eps) + 1 - (1 - d - O(\eps))^k$$
$$\leq O(k \eps) + 1 - ((1 - d)(1 - O(\eps))^k = 1 - (1 - d)^k + O(k \eps).$$
Given the setting of $\eps$, there is a very small constant $\delta$ such that this is:
$$\leq 1 - (1 - d)^k + \delta.$$
This implies that $X_0^{\otimes k}$ and $X_1^{\otimes k}$ are $(1 - (1 - d)^k  + \delta )$-indistinguishable for functions in $\mathcal{F}$.
\end{proof}

\paragraph{Comparing Theorems \ref{thm:geier} and \ref{thm:geier-implication}} We compare several aspects of the result proven in \cite{geier2022tight} and the similar result that our previous results imply. We begin by comparing the circuit sizes in the assumptions of the theorems. First, to obtain the stated indistinguishability against circuits of size $s$, we assume $d$-indistinguishability for functions in $\mathcal{F}''$, where $\mathcal{F}''$ consists of circuits of size $O(s \cdot \text{poly}(k) \cdot \log(|\mathcal{X}|))$, more specifically circuits of size $O(k^6 \log (k |\mathcal{X}|) + s k^2 + k^2 \log (|\mathcal{X}|))$. The bounds on the circuit sizes for the assumptions and implications of Theorem \ref{thm:geier} differ both qualitatively (in terms of how the different variables interact in the bounds) and Geier's result is stronger quantitatively. We did not focus on optimizing for the difference in the circuit sizes, and instead focus on showing how a comparable result follows immediately from our previous results.

Additionally, our results hold only for non-uniform circuits, while \cite{geier2022tight} also proves results for uniform circuits. It is an interesting problem for future work to formulate and prove a version of the multicalibration theorem that suffices for uniform complexity applications (similar to a uniform analog of the min-max theorem by \cite{vadhan2012characterizing}).

\section{Characterizing Distinguishability through Pseudo-Hellinger and Pseudo-R\'enyi Distance}\label{sec:pseudo-hellinger-renyi}

In this section, we discuss the connection between our distinguishability framework and the notion of pseudo-Hellinger distance. First, we provide a definition of pseudo-Hellinger distance with respect to general families of functions (instead of circuits of a certain size, as in the introduction).

\begin{definition}\label{def:mainPH}
    For random variables $X_0, X_1$ over $\mathcal{X}$, family $\mathcal{F}$ of functions, $\eps > 0$, the \emph{$(\mathcal{F}, \eps)$ pseudo-Hellinger distance} between $X_0$ and $X_1$ is the smallest $\delta$ such that there exist random variables $\widetilde{X}_0, \widetilde{X}_1$ such that:
    \begin{enumerate}
        \item $X_0$ is $\eps$-indistinguishable from $\widetilde{X}_0$ for functions $f \in \mathcal{F}$.
        \item $X_1$ is $\eps$-indistinguishable from $\widetilde{X}_1$ for functions $f \in \mathcal{F}$.
        \item $d_{\mathrm{H}}(\widetilde{X}_0, \widetilde{X}_1) = \delta$.
    \end{enumerate}
\end{definition}

With this definition in hand, we can derive the following characterization of the indistinguishability of $X_0^{\otimes k}, X_1^{\otimes k}$. Roughly speaking, this can be viewed as a generalization of Inequality~(\ref{ineq:hellinger}) to the efficient-distinguisher regime:

\begin{theorem}[Generalization of \cref{thm:pseudohellinger}]\label{thm:pseudohellinger-restatement}
    Suppose $X_0, X_1$ have $(\mathcal{F}, \eps)$ pseudo-Hellinger distance $\delta$. Then for every $k$:
    \begin{enumerate}
        \item  $X_0^{\otimes k}, X_1^{\otimes k}$ are $\sqrt{2k \delta^2} + 2k \eps$-indistinguishable for functions $f \in \mathcal{F}^{\otimes k}$.
        \item $X_0^{\otimes k}, X_1^{\otimes k}$ are $(1 - e^{-k\delta^2} - 2k\eps)$ distinguishable by functions $f \in \mathcal{F}_{K_2,K_1}$, where $K_1 = O(\frac{k}{\eps^{12}} \cdot \log(|\mathcal{X}|) \cdot \log(|\mathcal{X}|/\eps)) + O(k \cdot \log(|\mathcal{X}|) \cdot \polylog(k / \eps))$ and $K_2 = O(k/\eps^6)$.
    \end{enumerate}
\end{theorem}

As an immediate consequence, observe that the number of samples needed to distinguish $X_0, X_1$ with constant advantage is $\Theta(1 / \delta^2)$, where $\delta$ is the pseudo-Hellinger distance. Now, we prove the theorem:

\begin{proof}
    To prove Part (1) we first observe that, by Parts (1) and (2) of \cref{def:mainPH} and a simple hybrid argument, the $(\mathcal{F}, \eps)$ indistinguishability of $X_0^{\otimes k}$ and $X_1^{\otimes k}$ is bounded above by the computational indistinguishability of $\widetilde{X}_0^{\otimes k}$ and $\widetilde{X}_1^{\otimes k}$, plus $2 k \eps$. Next, the computational indistinguishability of $\widetilde{X}_0^{\otimes k}$ and $\widetilde{X}_1^{\otimes k}$ is upper-bounded by the total variation distance between $\widetilde{X}_0^{\otimes k}, \widetilde{X}_1^{\otimes k}$, which by Inequality~(\ref{ineq:hellinger}) is bounded above by $\sqrt{2 k d_H^2 \left(\widetilde{X}_0, \widetilde{X}_1\right)} = \sqrt{2 k \delta^2}$. This gives us the bound in Part (1) of the theorem.

To prove Part (2), we 
use the guarantees of \cref{thm:mainRestated} to obtain $\widetilde{X}_0, \widetilde{X}_1$ that are $\eps$-indistinguishable from $X_0,X_1$ such that the computational distinguishability of $X_0^{\otimes k}, X_1^{\otimes k}$ by $f \in \mathcal{F}_{K_2, K_1}$ is at least $d_{\mathrm{TV}}(\widetilde{X}_0^{\otimes k}, \widetilde{X}_1^{\otimes k}) - 2 k \eps$. In turn, Inequality~(\ref{ineq:hellinger}) tells us that this is at least $1 - e^{-k \cdot d^2_{\mathrm{H}}(\widetilde{X}_0, \widetilde{X}_1)} - 2 k \eps$, which is then at least $ 1 - e^{-k \delta^2} -2k \eps$, where $\delta$ is the pseudo-Hellinger distance, which is the smallest $d_{\mathrm{H}}(\widetilde{X}_0, \widetilde{X}_1)$ over all $\widetilde{X}_0, \widetilde{X}_1$ that are $\eps$-indistinguishable from $X_0, X_1$. This yields the theorem, as we desire.
\end{proof}

We now define pseudo-R\'enyi $1/2$-entropy with respect to general families of functions (instead of with respect to circuits of a certain size, as in the introduction).

\begin{definition}\label{def:pseudorenyi-restatement}
    For random variable $X_0$ over $\mathcal{X}$, family $\mathcal{F}$ of functions, $\eps > 0$, the \emph{$(\mathcal{F}, \eps)$ pseudo-R\'enyi $\frac{1}{2}$-entropy} of $X_0$ is the smallest $r$ such that there exists a random variable $\widetilde{X}_0$ such that:
    \begin{enumerate}
        \item $X_0$ is $\eps$-indistinguishable from $\widetilde{X}_0$ for functions $f \in \mathcal{F}$.
        \item $H_{1/2}\left(\widetilde{X}_0\right) = r$.
    \end{enumerate}
\end{definition}

We prove the following characterization of the indistinguishability of $X_0^{\otimes k}$ from the uniform distribution over $\mathcal{X}^{k}$.

\begin{theorem}[Restatement of \Cref{thm:pseudorenyi}]
   Suppose $X_0$ is a distribution over $\mathcal{X}$ that has $(\mathcal{F}, \eps)$ pseudo-R\'enyi $\frac{1}{2}$-entropy $r = \log|\mathcal{X}| - \delta$, where $\mathcal{F}$ is a family of functions that is closed under marginals. Let $\mathcal{U}$ be the uniform distribution over $\mathcal{X}$. Then for every $k$:
    \begin{enumerate}
        \item  $X_0^{\otimes k}, \mathcal{U}^{\otimes k}$ are $O(\sqrt{k \delta}) + 2k \eps$-indistinguishable for functions $f \in \mathcal{F}.$
        \item $X_0^{\otimes k}, \mathcal{U}^{\otimes k}$ are $(1 - e^{-\Omega(k \min\{\delta, 1\})} - 2k\eps)$ distinguishable by functions $f \in \mathcal{F'}_{K_2, K_1}$
    with $K_1 = O(k \cdot \log(|\mathcal{X}| / \eps) \cdot \log(|\mathcal{X}|)/ \eps^{24} ) + O(k \log(|\mathcal{X}|) \cdot \mathrm{polylog}(1 / \eps))$, and $K_2 = O(k / \eps^{12})$.
    \end{enumerate}
\end{theorem}

\begin{proof}
    To prove Part (1), following the same steps as in the proof of \Cref{thm:pseudohellinger-restatement}, by Definition \ref{def:pseudorenyi} and a hybrid argument we can bound the computational indistinguishability of $X_0^{\otimes k}$ and $\mathcal{U}^{\otimes k}$ by the computational indistinguishability of $\widetilde{X}_0^{\otimes k}$ and $\mathcal{U}^{\otimes k}$, plus $2 k \eps$. This is upper-bounded by the total variation distance between $\widetilde{X}_0^{\otimes k}$ and $\mathcal{U}^{\otimes k}$, plus $2 k \eps$, which is itself bounded above by $\sqrt{2k d_H^2\left(\widetilde{X}_0, \mathcal{U}\right)} + 2 k \eps$. The bound in Part (1) of the theorem is then obtained by applying \Cref{renyi-and-hellinger}, which tells us $d_H^2\left(\widetilde{X}_0, \mathcal{U}\right) = 1 - \sqrt{2^{- \delta}} \leq \sqrt{\delta}$.

    To prove Part (2), again as in the proof of \Cref{thm:pseudohellinger-restatement}, use \cref{thm:main} (more specifically, its generalization in \cref{thm:oneIntermediateMain}) to obtain a $\widetilde{X}_0$ that is $\eps$-indistinguishable from $X_0$ such that the computational distinguishability of $X_0^{\otimes k}, \mathcal{U}^{\otimes k}$ by functions $f \in \mathcal{F}_{K_2, K_1}$ is at least $d_{\mathrm{TV}}\left(\widetilde{X}_0^{\otimes k}, \mathcal{U}^{\otimes k}\right) - 2 k \eps$. As before, this is at least $1 - e^{-k \cdot d^2_{\mathrm{H}}(\widetilde{X}_0, \mathcal{U})} - 2 k \eps$, which can be related to the R\'enyi entropy of order $1/2$ of $\widetilde{X}_0$ using \Cref{renyi-and-hellinger}. \Cref{renyi-and-hellinger} tells us $d_H^2\left(\widetilde{X}_0, \mathcal{U}\right) = 1 - \sqrt{2^{- \delta}}$, which is $\Omega(\min\{1, \delta\})$, giving the theorem. 
\end{proof}

\section{General Product Distributions}\label{section:general-product-distributions}

In this section, we demonstrate that our results and proof techniques generalize to the setting of distinguishing general product distributions. Suppose we want to study indistinguishability with respect to $\mathcal{F}$. As before, we then define intermediate variables with respect to some $\mathcal{F}'$ such that $\mathcal{F}_{c \log |\mathcal{X}|, c} \subseteq \mathcal{F}'$. For example, if $\mathcal{F}$ is the family of functions given by circuits of size $\leq s$, then $\mathcal{F}'$ is the family of functions corresponding to circuits of size $s \cdot c + c \cdot \log |\mathcal{X}|$.

In what follows, consider $k$ pairs of random variables $\{(Y^a_0, Y^a_1)\}_{a = 1}^k$. Define $\{(\widetilde{Y}^a_0, \widetilde{Y}^a_1)\}_{a = 1}^k$ to be the corresponding random variables obtained by the guarantees of \cref{thm:MCproperties_restatement} applied for each pair $(Y^a_0, Y^a_1)$. 
Let $Y^1_0 \otimes \dots \otimes Y^k_0$ be the random variable obtained by concatenating all random variables $Y^a_0$; define $Y^1_1 \otimes \dots \otimes Y^k_1$ similarly.

We prove that the distinguishability of the two product distributions is characterized by the statistical distance of $\widetilde{Y}^1_0 \otimes \dots \otimes \widetilde{Y}^k_0$ and $\widetilde{Y}^1_1 \otimes \dots \otimes \widetilde{Y}^k_1$, paralleling the results of the previous sections. Specifically, we prove the following.

\begin{theorem}\label{thm:main-general-product}
    For every set of pairs of random variables $\{(Y_0^j, Y_1^j)\}_{j = 1}^k$, and family $\mathcal{F}$ of functions $f:\mathcal{X} \to \{0,1\}$ that is closed under marginals, and every $\eps>0$, there exist a set of pairs of random variables $\{(\widetilde{Y}_0^j, \widetilde{Y}_1^j)\}_{j = 1}^k$ such that for every $k > 0$,
    \begin{enumerate}
        \item $Y_b^j$ is $\eps$-indistinguishable from $\widetilde{Y}_b^j$ by functions in $\mathcal{F}$, for each $b\in \zo$ and $j \in [k]$.
        \item $Y^1_0 \otimes \dots \otimes Y^k_0 $ and $Y^1_1 \otimes \dots \otimes Y^k_1$ are $(d_{\mathrm{TV}}(\widetilde{Y}^1_0 \otimes \dots \otimes \widetilde{Y}^k_0, \widetilde{Y}^1_1 \otimes \dots \otimes \widetilde{Y}^k_1)+2k \cdot \eps)$-indistinguishable by functions in $\mathcal{F}$.
        \item $Y^1_0 \otimes \dots \otimes Y^k_0 $ and $Y^1_1 \otimes \dots \otimes Y^k_1$ are $(d_{\mathrm{TV}}(\widetilde{Y}^1_0 \otimes \dots \otimes \widetilde{Y}^k_0, \widetilde{Y}^1_1 \otimes \dots \otimes \widetilde{Y}^k_1)- 2k \cdot \eps)$-distinguishable by functions in $\mathcal{F'}_{K_2,K_1}$, where $K_1 = O(\frac{k}{\eps^{12}} \cdot \log(|\mathcal{X}|) \cdot \log(|\mathcal{X}|/\eps)) + O(k \cdot \log(|\mathcal{X}|) \cdot \polylog(k / \eps))$ and $K_2 = O(k/\eps^6)$.
    \end{enumerate}
\end{theorem}

By definition of $(\widetilde{Y}_0^j, \widetilde{Y}_1^j)$ with respect to $(Y_0^j, Y_1^j)$, \Cref{thm:main-general-product} follows immediately from \cref{thm:MCproperties_restatement}. We prove parts 2 and 3 of the theorem in the next two subsections.

\subsection{Proof of \cref{thm:main-general-product} Part (2)}

    We first make note of the following generalization of \Cref{lem:moving-to-product-distributions}.

\begin{lemma}\label{lemma:moving-to-product-distributions-general}
    Let $\mathcal{F}$ be a family of functions mapping from $\mathcal{X} \rightarrow [0,1]$. 
    For family $\mathcal{F}$ and each pair of random variables $Y_0^a, Y_1^a$ for $a \in [k]$, consider the construction of random variables $\widetilde{Y}_0^a, \widetilde{Y}_1^a$ as promised by \cref{thm:MCproperties_restatement}. Consider $\mathcal{F}^{\otimes k}$ as in \cref{def:marginals}. Then, for $f \in \mathcal{F}^{\otimes k}$ and $i \in \{0, 1\}$, 
    $$\left| \sum_{\Bar{x} \in \mathcal{X}^{k}} f(\Bar{x}) \cdot \left(\mathbb{P}[Y^1_i Y^2_i \dots Y^k_i = \Bar{x}] - \mathbb{P}[\widetilde{Y}^1_i \widetilde{Y}^2_i \dots \widetilde{Y}^k_i = \Bar{x}] \right) \right| \leq O(k \eps).$$
\end{lemma}

\begin{proof}
    The proof follows using the exact same steps as the proof of \cref{lem:moving-to-product-distributions}, replacing each $X_{b}^{\otimes (k-j)} \widetilde{X}_{i}^{j}$ for $b \in \{0, 1\}$ and $j \in [k]$ with $Y_b^1  Y_0^2 \dots Y_b^{(k-j)} \widetilde{Y}_b^{(k-j + 1)} \widetilde{Y}_b^{(k-j + 1)} \dots \widetilde{Y}_b^{k}$.
\end{proof}

Note that if $\mathcal{F}$ is a family of circuits of size $s$ for some $s \in \mathbb{N}$, it will be closed under marginals.

As in previous sections, we make note of the connection between the computational and statistical indistinguishability of $\widetilde{Y}^1_0 \otimes \dots \otimes \widetilde{Y}^k_0$ and  $\widetilde{Y}^1_1 \otimes \dots \otimes \widetilde{Y}^k_1$.

\begin{remark}\label{remark:computational_statistical_general}
    Let $\mathcal{F}$ be a family of functions mapping from $\mathcal{X} \rightarrow [0,1]$. 
    For family $\mathcal{F}$ and each pair of random variables $Y_0^a, Y_1^a$ for $a \in [k]$, consider the random variables $\widetilde{Y}_0^a, \widetilde{Y}_1^a$ as promised by \cref{thm:MCproperties_restatement}. Consider $\mathcal{F}^{\otimes k}$ as in \cref{def:marginals}. 
 For $f \in \mathcal{F}^{\otimes k}$, the following inequality holds: $$\left|\sum_{\Bar{P} \in \mathcal{P}^{\otimes k}} \sum_{\Bar{x} \in \Bar{P}} f(x) \cdot \left( \mathbb{P}[\widetilde{Y}^1_0 \otimes \dots \otimes \widetilde{Y}^k_0 = \Bar{x}] - \mathbb{P}[\widetilde{Y}^1_1 \otimes \dots \otimes \widetilde{Y}^k_1 = \Bar{x}]\right) \right|$$$$ \leq d_{\mathrm{TV}}(\widetilde{Y}^1_0 \otimes \dots \otimes \widetilde{Y}^k_0, \widetilde{Y}^1_1 \otimes \dots \otimes \widetilde{Y}^k_1).$$
\end{remark}

We can now prove the following lemma.

\begin{lemma}\label{lemma:computational-indistinguishability-bound-general}[\Cref{thm:main-general-product} Part (2)]
Let $\mathcal{F}$ be a family of functions mapping from $\mathcal{X} \rightarrow [0,1]$. 
    For family $\mathcal{F}$ and each pair of random variables $Y_0^a, Y_1^a$ for $a \in [k]$, consider the random variables $\widetilde{Y}_0^a, \widetilde{Y}_1^a$ as promised by \cref{thm:MCproperties_restatement}. Consider $\mathcal{F}^{\otimes k}$ as in \cref{def:marginals}. For $f \in \mathcal{F}^{\otimes k}$ the following inequality holds:
$$\left| \sum_{\Bar{x} \in \mathcal{X}^{k}} f(\Bar{x}) \cdot \left(\mathbb{P}[Y^1_0 \otimes \dots \otimes Y^k_0 = x] - \mathbb{P}[Y^1_1 \otimes \dots \otimes Y^k_1 = x] \right) \right| $$ $$\leq O(k \eps) + d_{\mathrm{TV}}(\widetilde{Y}^1_0 \otimes \dots \otimes \widetilde{Y}^k_0, \widetilde{Y}^1_1 \otimes \dots \otimes \widetilde{Y}^k_1).$$
\end{lemma}

\begin{proof}
    For $f \in \mathcal{F}^{\otimes k}$,
$$\left| \sum_{\Bar{x} \in \mathcal{X}^{k}} f(\Bar{x}) \cdot \left(\mathbb{P}[Y^1_0 \otimes \dots \otimes Y^k_0 = x] - \mathbb{P}[Y^1_1 \otimes \dots \otimes Y^k_1 = x] \right) \right| $$ $$\leq \left| \sum_{\Bar{x} \in \mathcal{X}^{k}} f(\Bar{x}) \cdot \left(\mathbb{P}[Y^1_0 \otimes \dots \otimes Y^k_0 = \Bar{x}] - \mathbb{P}[\widetilde{Y}^1_0 \otimes \dots \otimes \widetilde{Y}^k_0 = \Bar{x}] \right) \right|$$
$$+ \left| \sum_{\Bar{x} \in \mathcal{X}^{k}} f(\Bar{x}) \cdot \left(\mathbb{P}[\widetilde{Y}^1_0 \otimes \dots \otimes \widetilde{Y}^k_0 = \Bar{x}] - \mathbb{P}[\widetilde{Y}^1_1 \otimes \dots \otimes \widetilde{Y}^k_1 = \Bar{x}] \right) \right| $$ $$+ \left| \sum_{\Bar{x} \in \mathcal{X}^{k}} f(\Bar{x}) \cdot \left(\mathbb{P}[\widetilde{Y}^1_1 \otimes \dots \otimes \widetilde{Y}^k_1 = \Bar{x}] - \mathbb{P}[Y^1_1 \otimes \dots \otimes Y^k_1  = \Bar{x}] \right) \right|.$$
From the previous sections and the remark above, this is:
$$\leq O(k \eps) + d_{\mathrm{TV}}(\widetilde{Y}^1_0 \otimes \dots \otimes \widetilde{Y}^k_0, \widetilde{Y}^1_1 \otimes \dots \otimes \widetilde{Y}^k_1). \eqno \qedhere$$
\end{proof}

\subsection{Proof of \cref{thm:main-general-product} Part (3)}

We prove the following theorem:

\begin{lemma}\label{thm:efficient-distinguishers-general}[\cref{thm:main-general-product} Part (3)]
    Let $Y^1_0 \otimes \dots \otimes Y_0^k, Y^1_1\otimes \dots \otimes Y_1^k$ be product distributions.
     Let $\mathcal{F}$ be a family of functions mapping from $\mathcal{X} \rightarrow [0,1]$. 
    For each pair of random variables $Y_0^a, Y_1^a$ for $a \in [k]$, consider the random variables $\widetilde{Y}_0^a, \widetilde{Y}_1^a$ as promised by \cref{thm:MCproperties_restatement} with respect to a family of functions $\mathcal{F}$. Then, there exists a function in 
    \[
     \mathcal{F}_{K_2, K_1},
    \]
    where $K_1 = O(\frac{k}{\eps^{12}} \cdot \log(|\mathcal{X}|) \cdot \log(|\mathcal{X}|/\eps)) + O(k \cdot \log(|\mathcal{X}|) \cdot \polylog(k / \eps))$ and $K_2 = O(k/\eps^6)$,
    which distinguishes $Y^1_0 \otimes \dots \otimes Y_0^k, Y^1_1\otimes \dots \otimes Y_1^k$ with advantage 
    \[
    d_{\mathrm{TV}}(\widetilde{Y}^1_0 \otimes \dots \otimes \widetilde{Y}_0^k, \widetilde{Y}^1_1\otimes \dots \otimes \widetilde{Y}_1^k) -4 \delta k.
    \]
\end{lemma}

\begin{proof}
    We follow the same proof procedure as in \cref{thm:distinguishingCircuit}. Observe that the only difference is that instead of re-using the same values of $\Pr[\widetilde{X}_0 \in P_i]$ and $\Pr[\widetilde{X}_1 \in P_i]$ for each sample, we instead use the probabilities with respect to the corresponding multicalibrated partition for each sample. I.e., we instead analyze $\Pr[\widetilde{Y}_0^a \in P_i^a]$ and $\Pr[\widetilde{Y}_1^a \in P_i^a]$, where now the partitions $P_i^a \in \mathcal{P}^a$, for $a \in [k]$.
Now, for such a part $P^a_i \in \mathcal{P}^a$ let 
\[
\alpha^{(a)}_0(P_i) = \frac{\Pr[\widetilde{Y}^a_0 \in P^a_i]}{\Pr[\widetilde{Y}^a_0 \in P^a_i] + \Pr[\widetilde{Y}^a_1 \in P^a_i]},
\]
and define $\alpha^{(a)}_1(P_i)$ similarly. 

Next, let us define new distributions $\widetilde{Y}^a_{0, \delta}, \widetilde{Y}^a_{1, \delta}$, where all of the $\alpha(P_i)$ values are arbitrarily shifted to a neighboring multiple of $\frac{\delta}{r}$. That is to say if $\alpha^{(a)}_0(P_i) \in [\ell \delta / r, (\ell+1)\delta / r]$ for an integer $\ell$, then we arbitrarily set $\alpha^{(a)}_{0, \delta}(P_i)$ to be one of $\ell \delta / r, (\ell+1)\delta / r$.

Observe that 
\[
d_{\mathrm{TV}}(\widetilde{Y}^a_{0,\delta}, \widetilde{Y}^a_0) \leq \delta,
\]
as there are $r$ parts in $\mathcal{P}$, and each has its probability mass shifted by at most $\delta / r$. Note that this also means that 
\[
d_{\mathrm{TV}}(\widetilde{Y}^1_{0, \delta} \otimes \dots \otimes \widetilde{Y}_{0, \delta}^k, \widetilde{Y}^1_0 \otimes \dots \otimes \widetilde{Y}_0^k) \leq \delta k.
\]

Now, our goal will be to construct an optimal distinguisher for $\widetilde{Y}^1_0 \otimes \dots \otimes \widetilde{Y}_0^k$ and $\widetilde{Y}^1_1 \otimes \dots \otimes \widetilde{Y}_1^k$. To do this, consider a sequence of samples $x_1, \dots x_k$, where $x_1 \in P_{i_1}, \dots , x_k \in P_{i_k}$. Observe that we can exactly compute the relative probability that $x_1, \dots x_k$ comes from $\widetilde{Y}^1_0 \otimes \dots \otimes \widetilde{Y}_0^k$ versus $\widetilde{Y}^1_1 \otimes \dots \otimes \widetilde{Y}_1^k$. 

Indeed, let 
\[
\kappa((x_1, \dots x_k)) = \frac{\Pr[\widetilde{Y}^1_{1, \delta} \otimes \dots \otimes \widetilde{Y}_{1, \delta}^k = (x_1, \dots x_k)]}{\Pr[\widetilde{Y}^1_{0, \delta} \otimes \dots \otimes \widetilde{Y}_{0, \delta}^k = (x_1, \dots x_k)]} = \prod_{a = 1}^k \frac{\alpha^{(a)}_{1, \delta}(P_{i_j})}{\alpha^{(a)}_{0, \delta}(P_{i_j})}.
\]

For a sequence of $k$ samples $x_1, \dots x_k$, the optimal distinguisher is simply the distinguisher which returns $1$ (corresponding to deciding that the samples are from $\widetilde{X}_{1, \delta}^{\otimes k}$) if the above expression is $\geq 1$, and $0$ otherwise (corresponding to deciding that the samples are from $\widetilde{X}_{0, \delta}^{\otimes k}$). This is because in general for two distributions $p, q$, we have that 
\[
d_{\mathrm{TV}}(p, q) = \sum_{x \in \mathcal{X}} (\max(p(x), q(x)) - \min(p(x), q(x))).
\]

That is to say, if we let $C((x_1, \dots x_k))$ be the circuit which computes $\mathbf{1}[\kappa((x_1, \dots x_k)) \geq 1]$, we have
\[
\left | \sum_{(x_1, \dots x_k) \in \mathcal{X}^k} C((x_1, \dots x_k)) \cdot \left [\widetilde{Y}^1_{0, \delta} \otimes \dots \otimes \widetilde{Y}_{0, \delta}^k(x_1, \dots x_k) - \widetilde{Y}^1_{1, \delta} \otimes \dots \otimes \widetilde{Y}_{1, \delta}^k(x_1, \dots x_k) \right ] \right | 
\]
\[
= d_{\mathrm{TV}}(\widetilde{Y}^1_{0, \delta} \otimes \dots \otimes \widetilde{Y}_{0, \delta}^k, \widetilde{Y}^1_{1, \delta} \otimes \dots \otimes \widetilde{Y}_{1, \delta}^k).
\]

Now, all that remains to show is that we can compute such a circuit $C$ which is not too complex. Indeed, the key observation is that because we ``rounded'' our distributions initially, now when we compute the product $\prod_{a = 1}^k \frac{\alpha^{(a)}_{1, \delta}(P_{i_j})}{\alpha^{(a)}_{0, \delta}(P_{i_j})}$, every pair of $\alpha^{(a)}_{1, \delta}(P_{i_j}), \alpha^{(a)}_{0, \delta}(P_{i_j})$ share a denominator. Thus, computing whether this product is $\geq 1$ is equivalent to multiplying together $k$ integers in $[r / \delta]$ and checking which is larger. So, we consider the following circuit: first, $C$ has $k$ copies of the multicalibration circuit which, upon receiving element $x_a$, computes the index $i$ for which $x_a \in P^{(a)}_i$ (defined with respect to the $a$th distribution). Recall that this circuit is $\in \mathcal{F}'_{t, q, r}$. Next, once we have computed the index $i$ of the partition that an element $x$ is in, we hard-wire two-lookup tables: one computes the numerator of $\alpha^{(a)}_{0, \delta}(P_i)$, and the other computes the numerator of $\alpha^{(a)}_{1, \delta}(P_i)$. Observe that because there are only $r$ partitions, the complexity of these look-up tables is not too large. By \cite{Lup70}, each look-up table requires only circuits of size $O(r)$. Once we have computed the numerators of these values, all that remains is to multiply together these integers, and return the index $b \in \zo$, corresponding to whichever product is larger. Computing this product can be done with circuits of size $O(k \cdot \polylog(r / \delta))$.

Thus, in total, this circuit $C$ for distinguishing $\widetilde{Y}^1_{0, \delta} \otimes \dots \otimes \widetilde{Y}_{0, \delta}^k, \widetilde{Y}^1_{1, \delta} \otimes \dots \otimes \widetilde{Y}_{1, \delta}^k$ can be implemented in \[\mathcal{F}_{qk, tk\log(|\mathcal{X}|) + O(rk) + O(k \cdot \polylog(r / \delta)), rk},
\]
as we use $k$ copies of a circuit in $\mathcal{F}_{t, q\log(|\mathcal{X}|)}$, along with $O(k)$ look-up tables, and $O(k \cdot \polylog(r / \delta))$ extra wires to compute the product. By plugging in $q = O(1 / \eps^6), r= O(1/\eps),t = O(1 /(\eps^{12} \gamma) \cdot \log(|\mathcal{X}|)/\eps )$, we obtain the stated complexity.

Finally, by the same manipulations as in conclusion of the proof of \cref{thm:distinguishingCircuit}, it will be the case that this circuit $C$ achieves distinguishing advantage at least $d_{\mathrm{TV}}(\widetilde{Y}^1_0 \otimes \dots \otimes \widetilde{Y}_0^k, \widetilde{Y}^1_1\otimes \dots \otimes \widetilde{Y}_1^k) - 4\delta k$ when applied to the distributions $Y_0^1 \otimes \dots \otimes Y_0^k, Y_1^1 \otimes \dots \otimes Y_1^k$. This concludes the proof.
\end{proof} 

Initializing the multicalibrated partitions with the same function class $\mathcal{F}$ as in \cref{lemma:moving-to-product-distributions-general} then yields a mirroring upper bound in terms of the size of distinguishing circuits. 

\section{Conclusion}

In this paper, we explored the problem of distinguishing product distributions through the lens of the Multicalibration Theorem. The main contribution of the paper was showing a near-equivalence between the computational indistinguishability of $X_0^{\otimes k}$ and $X_1^{\otimes k}$ and the statistical indistinguishability of $\widetilde{X}_0^{\otimes k}$ and $\widetilde{X}_1^{\otimes k}$, where $\widetilde{X}_0$ and $\widetilde{X}_1$ are efficiently computable given $X_0$ and $X_1$ through the guarantees of the Multicalibration Theorem. The number of samples to efficiently distinguish $X_0^{\otimes k}$ and $X_1^{\otimes k}$ is almost tightly characterized by $\Theta\left(d_H^{-2} \left(\widetilde{X}_0, \widetilde{X}_1 \right)\right)$, where $d_H$ is the Hellinger distance between random variables. Our paper demonstrates that multicalibration is a suitable lens through which to study the distinguishability of random variables due to its ability to transform computational notions of distinguishability into information-theoretic notions of distinguishability.

\subsection{Acknowledgements}

This project was inspired by a final project completed by the first two authors during Cynthia Dwork's course ``Topics in Theory for Society: The Theory of Algorithmic Fairness'' at Harvard (Spring 2024). We thank Cynthia Dwork for initial discussions regarding multicalibration and its applications to complexity and hardness amplification. We thank the anonymous Eurocrypt and CCC reviewers for their suggestions and feedback, and we thank Pranay Tankala for bringing to our attention errors in our cited version of the Multicalibration theorem.

\bibliographystyle{alpha}
\bibliography{ref}

\appendix

\section{Proof of \cref{cor:asymptotic}}\label{appendix:asymptotic-proof}

\begin{proof}[Proof of \cref{cor:asymptotic}]
    We start by proving Item 1 of \cref{cor:asymptotic}. By definition, if $X \equiv_{\delta}^{cH} Y$, this means that there exist ensembles $\widetilde{X} = \{\widetilde{X}_n \}, \widetilde{Y} = \{\widetilde{Y}_n\}$ and some function $\eta$ satisfying $\eta(n) = n^{-\omega(1)}$ such that $X \equiv^{\mathrm{comp}}_{\eta} \widetilde{X}$, $Y \equiv^{\mathrm{comp}}_{\eta} \widetilde{Y}$, and $\forall n$, $d_H(\widetilde{X}_n, \widetilde{Y}_n) \leq \delta(n)$. Further, using \Cref{def:computational-indistinguishability-ensembles}, this implies that there exists $s(n) = n^{\omega(1)}, \eps(n) = n^{-\omega(1)}$ such that for all $n$, $X_n$ is $(n^{-\omega(1)} + \eps(n))$-indistinguishable from $\widetilde{X}_n$ with respect to size $s(n)$ circuits (and analogously for $Y_n$ and $\widetilde{Y}_n$).

    Therefore, by a simple hybrid argument, $X_n^{k(n)}$ is $(k(n) \cdot (n^{-\omega(1)} + \eps(n)))$-indistinguishable from $\widetilde{X}_n^{k(n)}$ and $Y_n^{k(n)}$ is $(k(n) \cdot (n^{-\omega(1)} + \eps(n)))$-indistinguishable from $\widetilde{Y}_n^{k(n)}$, both with respect to size $s(n)$ circuits. 
    Additionally, by relating Hellinger distance to total variation distance, there is a choice of $k(n) = \Omega(1 / \delta(n)^2)$ such that  $\forall n$, $d_{\mathrm{TV}}(\widetilde{X}_n^{k(n)}, \widetilde{Y}_n^{k(n)}) \leq 1/4$. 

    The indistinguishability parameter (i.e. ``$\gamma$'' if two random variables are $\gamma$-indistinguishable) of $X^{k(n)}$ and $ Y^{k(n)}$ is upper bounded by the sum of the indistinguishability parameter of $X^{k(n)}, \widetilde{X}^{k(n)}$, of $\widetilde{X}^{k(n)}, \widetilde{Y}^{k(n)}$, and of $Y^{k(n)}, \widetilde{Y}^{k(n)}$. Therefore, $X^{k(n)}$ and $ Y^{k(n)}$ are $\left(\frac{1}{4} + k(n) \cdot (n^{-\omega(1)} + \eps(n))\right)$-indistinguishable, which is $\leq 1/2$ as $k(n) = n^{O(1)}$. Thus, $X \equiv^{cS}_k Y$.

    Now, we prove Item 2 of \cref{cor:asymptotic}. First, by definition, if $X \equiv_k^{cS} Y$, we have $X^k \equiv^{\mathrm{comp}}_{1/2} Y^k$. In particular, this means that there exists some $s'(n) = n^{\omega(1)}$ and $\eps'(n) = n^{-\omega(1)}$ such that 
    $X_n^{k(n)}$ and $Y_n^{k(n)}$ are  
    $(\frac{1}{2} + \eps'(n))$-indistinguishable by size $s'(n)$ circuits. Now, for all $n$, let us choose $s(n) = n^{\omega(1)}$ and $\eps(n) = n^{-\omega(1)}$ such that $s'(n) \geq O(s(n) k(n) / \eps^2) + \mathrm{poly}(k(n) / \eps)$, as in \cref{thm:pseudohellinger}. Next, if we set $\delta(n) = \frac{1000}{\sqrt{k(n)}}$, then $1/2 + \eps'(n) \leq 1 - e^{-k(n) \delta^2(n)} - 2 k(n) \eps(n)$ (recall that $k(n) \eps(n)$ is negligible), and thus $X_n^{k(n)}$ and $Y_n^{k(n)}$ are $(1 - e^{-k(n) \delta^2(n)} - 2 k(n) \eps(n))$-\emph{indistinguishable} by size $s'(n)$ circuits as well. 
    
    As an immediate consequence, when we invoke the contrapositive of Item 2 of \cref{thm:pseudohellinger}, we see that $X_n, Y_n$ have $(s(n), \eps(n))$ pseudo-Hellinger distance \emph{at most} $\delta(n)$. This yields the desired claim, as $\delta(n) = O \left ( \frac{1}{\sqrt{k(n)}}\right )$.
\end{proof}

\end{document}